\documentclass[aps,prl,twocolumn,showpacs,superscriptaddress,floatfix]{revtex4-1}
\usepackage[utf8]{inputenc}
\usepackage{ucs}
\usepackage{natbib}
\usepackage{graphicx}
\usepackage{bm}
\usepackage{amssymb}
\usepackage{amsmath}
\usepackage{xcolor}
\usepackage{upgreek}
\usepackage{braket}
\usepackage{soul}
\usepackage{hyperref}
\hypersetup{
    colorlinks=true,
    linkcolor=blue,
    filecolor=magenta,
    urlcolor=cyan,
    pdftitle={Overleaf Example},
    pdfpagemode=FullScreen,
    }

\def\cm{cm$^{-1}$}

\draft 

\begin{document}

\title{Novel Dipole-Lattice coupling in the Quantum-Spin-Liquid Material $\upkappa$-(BEDT-TTF)$_2$Cu$_2$(CN)$_3$}

\author{Jesse Liebman}

\affiliation{The Department of Physics and Astronomy, The Johns Hopkins University, Baltimore, Maryland 21218, USA}
\author{Kazuya Miyagawa}

\affiliation{University of Tokyo, Hongo, Tokyo, Japan}
\author{Kazushi Kanoda}

\affiliation{University of Tokyo, Hongo, Tokyo, Japan}

\author{Natalia Drichko}
\email[]{drichko@jhu.edu}

\affiliation{The Department of Physics and Astronomy, The Johns Hopkins University, Baltimore, Maryland 21218, USA}

\date{\today}

\begin{abstract}

A family of molecular Mott insulators on triangular lattice provided a few S=1/2 triangular quantum spin liquid candidates, with $\upkappa$-(BEDT-TTF)$_2$Cu$_2$(CN)$_3$ being the most studied material of this group. The large number experimental works present a conflicting set of evidence, with some suggesting spin liquid behavior, while others point towards  a valence bond solid with orphan spins. In this work we use Raman scattering spectroscopy to probe both local charge on molecular sites and lattice phonons as a function of temperature down to 6~K. Based on the analysis of the line shape of the BEDT-TTF charge sensitive vibration  $\nu_2$ on cooling below  40 K, we suggest a development of disordered fluctuating charge disproportionation on (BEDT-TTF)$_2$ dimers of amplitude as small as 0.06$e$. The lattice phonons show strong anomalous broadening on cooling only in the (c,c) scattering channel, associated with the developing charge disproportionation. We suggest an interpretation, where the coupling of disordered charge dipoles on dimers to the lattice results in anisotropic modulation of charge transfer integrals between dimer lattice sites. Such fluctuations would result in modulation of  magnetic coupling between spins which can produce fluctuating charge ordered spin-singlet pairs.

\end{abstract}

\maketitle

\section{Introduction}

In strongly correlated systems, exotic physics and critical behavior emerge at the junction of geometric frustration and competing interactions. One example is the quantum spin liquid (QSL), where emergent fractional spin excitations arise from the infinitely-degenerate collective electronic quantum state. Ever since Anderson suggested a Resonating Valence Bond (RVB) ground state~\cite{Anderson1973}, Mott insulators with spins $S=\frac{1}{2}$ on a triangular lattice with nearest-neighbor anitferromagnetic interactions have been studied in hopes of realizing the QSL. However, further studies have shown that realizing the QSL requires interactions beyond nearest neighbors~\cite{Huse1988, Misguich1999, Zhu2015}. 
Molecular spin liquid candidates, one of which we discuss in the present work, were suggested as providing an alternative way for competing interactions, involving charge degree of freedom in addition to magnetic one~\cite{Hotta2010}. The discussion if these degrees of freedom would compete or interact, resulting in  a new type of spin liquid is still ongoing.  

\begin{figure}[h]
\centering
\includegraphics[width=\linewidth]{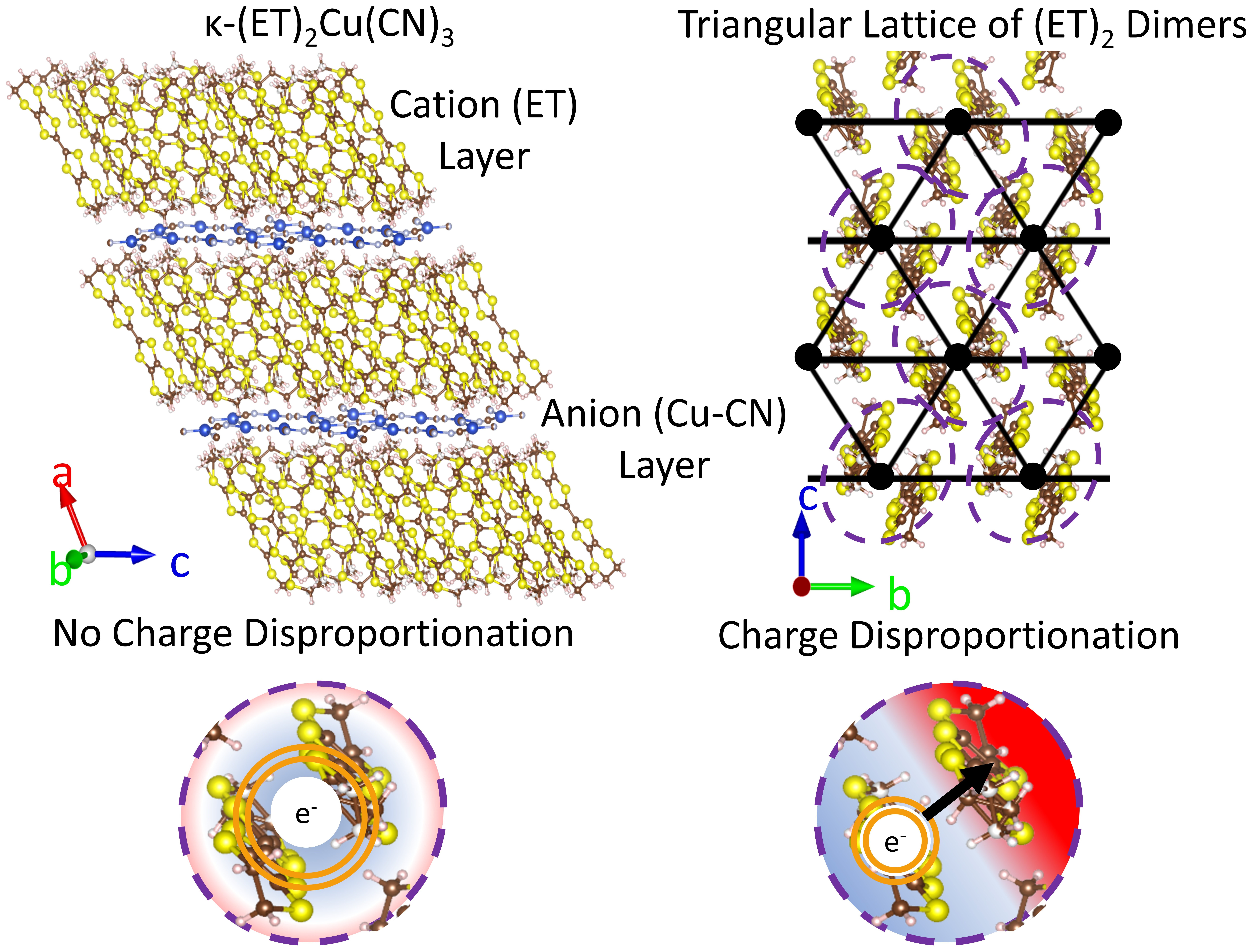}
\caption{\label{Structure} Structure of $\upkappa$-(ET)$_2$Cu$_2$(CN)$_3$. ET layers are separated by Cu-CN layers. ET molecules form dimers on a triangular lattice. Within an (ET)$_2$ dimer, the HOMO electron can be shared equally (no charge disproportionationer left panel) or unequally (charge disproportionation lower right panel) with the resulting Electric Dipole vector shown as a black arrow.}
\end{figure}

Molecular Mott insulators with so-called $\upkappa$-phase structure can realize  $S=\frac{1}{2}$ QSL candidates on a triangular lattice. In these layered molecular materials the lattice of the functional layer is formed by pairs of ET molecules~\footnote{ET refers to bis(ethylenedithio)tetrathiafuvalene} that share one hole (+1$e$). These dimer ET$_2^{1+}$ spin $S=1/2$ units form a crystal layer, where overlapping $\pi$-orbitals of ET molecules result in electronic bands. The large ratio $U/t$ of the electronic overlap between the dimers $t$ to on-site electron repulsion $U$ results in a a Mott insulator state for some of these materials, with anitferrmagnetic interaction between S=1/2. 
Two possible scenarios are proposed for these materials to realize QSL state, one is a ring exchange preventing 120 degree order ~\cite{Misguich1999, Holt2014}. Another is an exotic ``quantum dipole liquid''~\cite{Hotta2010}, where spins are coupled to the fluctuating electric dipoles formed on molecular (ET)$_2$ as a result of electronic correlations (see Fig.~\ref{Structure} for the illustration of the charge degree of freedom). The quantum dipole liquid, first suggested in connection with the non-trivial dielectric behavior of $\upkappa$-(ET)$_2$Cu$_2$(CN)$_3$ ~\cite{Abdel2010,Pinteric2014,Yakushi2015}, was observed in another molecular Mott insulator~\cite{Hassan2018}.

$\upkappa$-(ET)$_2$Cu$_2$(CN)$_3$  ($\upkappa$-Cu-CN) is the most thoroughly studied  of these materials.  Spins in the magnetic lattice of $\upkappa$-Cu-CN experience antiferromagnetic (AF) interactions  $J$  on the order of 200 K~\cite{Shimizu2003}. The QSL ground state of this material was suggested based on the lack of magnetic ordering observed down to 32~mK~\cite{Shimizu2003}, despite large AF interactions. Robustness of a description by S=1/2 magnetism on a triangular lattice with $J$ of about 200~K was confirmed by observation of spin excitations in Raman scattering~\cite{Nakamura2014,Hassan2018}. A linear contribution to the specific heat that is insensitive to magnetic field  was attributed to gapless fermionic excitations~\cite{Yamashita2008}.

The QSL nature of $\upkappa$-Cu-CN has been actively discussed recently, starting with the  thermal transport~\cite{Yamashita2009} measurements.
The nuclear magnetic resonance (NMR)~\cite{Shimizu2006}  and electron spin resonance (ESR)~\cite{Miksch2021} measurements indicate a low-temperature decrease in spin susceptibility reminiscent of spin-singlet formation with a finite residual susceptibility, which was argued to come from orphan spins~\cite{Riedl2019,Miksch2021,Pustogow2020}. Multiple spectroscopic studies of $\upkappa$-Cu-CN  material~\cite{Elasser2012,Yakushi2015,Sedlmeier2012,Dressel2016,Matsuura2022} pointed on possible unconventional behavior of the charge degree of freedom and the lattice, but have not produced conclusive results.

Presented in this work Raman scattering measurements of $\upkappa$-Cu-CN conducted  in a broad frequency range from 12 up to 2000~\cm\ down to 6 K allowed us to uncover glassy (disordered) charge fluctuations on the ET$_2$ dimers and their interaction with the lattice. The quantitative analysis of the line shapes of lattice and  molecular vibrations allow us  to follow the temperature dependence of the frequency of charge fluctuations $\omega_{\text{decay}}$. Our observations are consistent with the emergence of a charge dipole glass state in $\upkappa$-Cu-CN. We speculate that the dipole glass state corresponds to disorder in the magnetic exchange constant $J$, which promotes spin-singlet formation and a valence bond glass state.

\section{Results}

\begin{figure}[h]
\includegraphics[width=\linewidth]{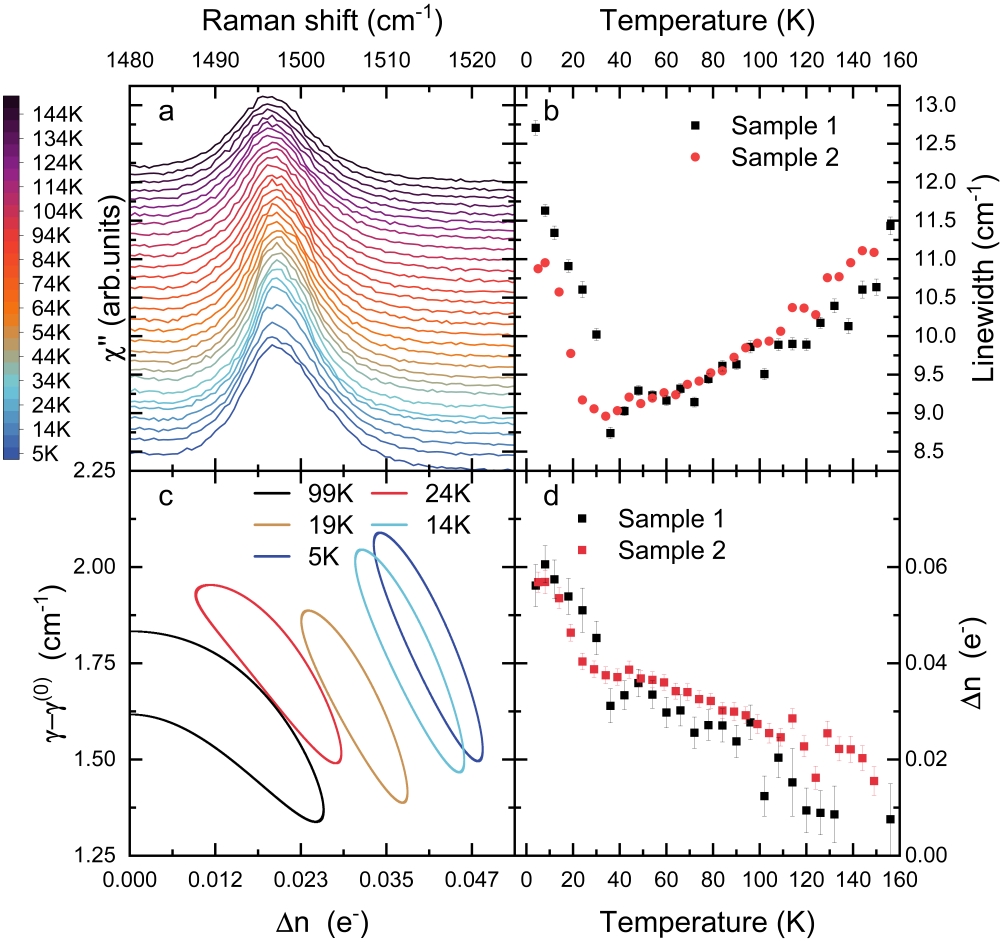}
\caption{\label{Mol Vib}{\bf (a)} Raman scattering spectra of $\upkappa$-(ET)$_2$Cu$_2$(CN)$_3$ in the frequency region of molecular vibration $\nu_2$ in the (b,c) scattering channel,  Sample 2. Spectra are shifted along y axis for clarity. {\bf (b)} The temperature dependence of the linewidth of $\nu_2$ molecular vibration. {\bf (c)}  Boundaries of 1$\sigma$ confidence regions for fit parameters of the Charge Glass Model for sample Sample 2: charge disproportionation $\Delta n$ and Lorentzian width parameter $\gamma-\gamma^{(0)} \propto \omega_{\text{decay}}^{-1}$, where $\omega_{\text{decay}}$ is charge fluctuation frequency. Note the narrowing of the parameters region and the increase of $\Delta n$ with temperature decrease.  {\bf (d)} Temperature dependence of the charge disproportionation $\Delta n$ obtained from  the fit of the $\nu_2$ line shape with the case of static charge glass model. 
}
\end{figure}

As the first step, we re-examine the line shape of a vibration of the central C=C bond of ET molecules $\nu_2$, which is an accepted tool to study the distribution of the charge on the Highest Occupied Molecular Orbital (HOMO) of an ET molecule. For a detailed explanation of the line shape analysis and fitting procedure, see {\color{red} SI}. The  frequency of $\nu_2$ (in \cm) satisfies the relation
\begin{equation}\label{Eq 0}
\omega(n) = 1447 \text{~\cm} + 120(1e^--n)\frac{\text{\cm}}{e^-}
\end{equation}
where $n$ is the average charge on the ET molecule, with $n=0.5e^-$ in most ET-based material~\cite{Yamamoto2005}. As shown in the Fig.~\ref{Mol Vib} a, we observe that $\nu_2$ narrows on cooling down to approximately 40 K, but broadens on cooling below this temperature, as estimated by a temperature dependence of the Full-Width-Half-Maximum (FWHM) (see Fig.~\ref{Mol Vib}b). In Fig.~\ref{Mol Vib} we show results obtained for two samples, which show a similar behaviour with small discrepancies in temperature behavior and linewidth. Its frequency does not show any anomaly or splitting upon cooling, in agreement with the previous studies~\cite{Yakushi2015}.
In our analysis of the line shape of $\nu_2$, we considered two possible scenarios:

(i)  Static or fluctuating charge order between sites A and B with a well-defined charge separation, $\Delta n=n_A-n_B$, which results in a frequency splitting, $\Delta\omega_{AB}$. Charge disproportionation within an (ET)$_2$ dimer creates an electric dipole, as shown in Fig.~\ref{Structure}. We assume each component possesses a similar life time as the non-perturbed $\nu_3$ vibration, which provides ``natural'' linewidth (in our case, 5~\cm). A frequency of fluctuations, $\omega_{ex}$, is defined as the hopping rate between sites~\cite{Kubo1969,Sue1986,Yakushi2012_Crystals, Hassan2018}. Static charge order is the case with $\omega_{ex}$=0. The static and dynamic cases result in a renormalized Lorentzian lineshape of the response.

(ii) Charge glass: A Gaussian parameter describes a random distribution of charge by charge fluctuations with a stochastic decay process. The charge glass is modeled by the autocorrelation function of charge fluctuations
\begin{equation}\label{Eq 1}
\Delta^2_{n}(\tau) \equiv \langle \Delta n (\tau) \Delta n (0) \rangle = \Delta^2_{n} \exp(-\omega_{\text{decay}}\tau)
\end{equation}
where $\Delta_{n}$ is the amplitude of charge fluctuations and $\omega_{\text{decay}}$ is the inverse lifetime of charge fluctuations~\cite{Kubo1969, Sue1986, Yakushi2012_Crystals}. 
In the dimerized $\upkappa$-phase, we expect in general, that for the A and B molecules of a dimer, $n_A+n_b=+1e$. We can use the simplified version of the Eq.~\ref{Eq 1}, since this rule imposed by dimerization only influences the amplitude the charge separation $\Delta n$, which is an experimentally fit parameter.
The resulting lineshape is a renormalized Voigt profile, characterized by a Lorentzian width $\gamma$ determined by decay frequency $\omega_{\text{decay}}$.
\begin{equation}\label{Eq 2}
\gamma = \gamma_0 + C^2\frac{\Delta \omega^2}{\omega_{\text{decay}}}
\end{equation}
and Gaussian variance
\begin{equation}\label{Eq 3}
\sigma^2 = \sigma_{\text{res}}^2+\Delta\omega^2
\end{equation}
where $\gamma_0$ is the natural lifetime of the pure vibration, $C$ is a constant, $\Delta\omega = \frac{120~\text{\cm}}{e^-}\Delta_n$, and $\sigma_{\text{res}}$ is the spectral resolution. 
Following this model, as the static limit is approached, the Lorentzian width diverges as $\omega_{\text{decay}} \to 0$. Once the dynamic process freezes out entirely $(\omega_{\text{decay}} = 0)$, the Lorentzian width renormalizes to $\gamma_0$. 
In a static case the Gaussian variance determined by $\Delta\omega$ will define the line shape. 

For the case of a relatively weak broadening of $\nu_2$, the analysis of the experimental spectra with both models has the following limitation: we cannot uniquely determine the charge separation and fluctuation rate, which may be both temperature dependent, and have large covariance.
The increase of charge separation  leads to an increase of the the linewidth (Lorenzian for model (i) and Gaussian for model (ii)), while the increase of fluctuation frequency leads to a decrease of the Lorentzian linewidth.

A static case of both models provides a lower limit estimate of the charge separation. A fit with the static charge order (model (i))  yields $\Delta n$ = 0.05$e$ at the lowest measured temperature of 6~K, which is shown in {\color{red}SI}. 
The static case ($\omega_{\text{decay}} \to 0$) of the charge glass (model (ii)) yields a standard deviation of charge disproportionation of 0.03$e$, corresponding to an average charge separation of 0.05e in a dimer.  Fig. 2 d shows the temperature dependence of the average charge separation for a static case. However, the static case provides worse fit quality than dynamic charge disproportionation.

To identify the best fit parameters we used the total squared error as a measure of goodness of fit (see {\color{red}{SI}}
) calculated as $TSE_{\text{model}} = \sum_{i=1}^N(I(\omega)_{\text{meas}}-I(\omega)_{\text{model}})^2$. While the error is quantitatively different for the two measured samples, the final result is that the best fit is provided by the Gaussian shape (charge glass model).
The values of $\Delta n$ and $\omega_{\text{decay}}$ suggested by 95\% fit confidence regions~\cite{Efron1994} are shown in Fig.~\ref{Mol Vib}{\bf c}. The range of parameters narrows on cooling; Charge disproportionation shows an increase. The best fit provides a broad range for fluctuation frequencies with a temperature variation of frequency smaller than this range, as shown by the  95\% confidence regions at few different temperatures (see Fig. 2{\bf c}).

The main result of this analysis is that the charge disproportionation  abruptly increases below 40~K and begins to flatten near 20~K, remaining flat below 10~K, in agreement with the results of Ref.~\cite{Yamamoto2005,Sedlmeier2012}. We find that this charge disproportionation is disordered and slowly fluctuating. Vibrational analysis cannot follow the frequency of fluctuations precisely. The average charge separation at  the lowest temperatures measured about 6~K, is estimated to be 0.06e, which is  very close to the estimate provided by dielectric measurements in Ref.~\cite{Abdel2010}. This is significantly smaller than the charge separation observed in the only  material  $\upkappa$-ET$_2$Hg(SCN)$_2$Cl with static charge ordered state~\cite{Drichko2014,Hassan2020}.

\begin{figure}
    \includegraphics[width=\linewidth]{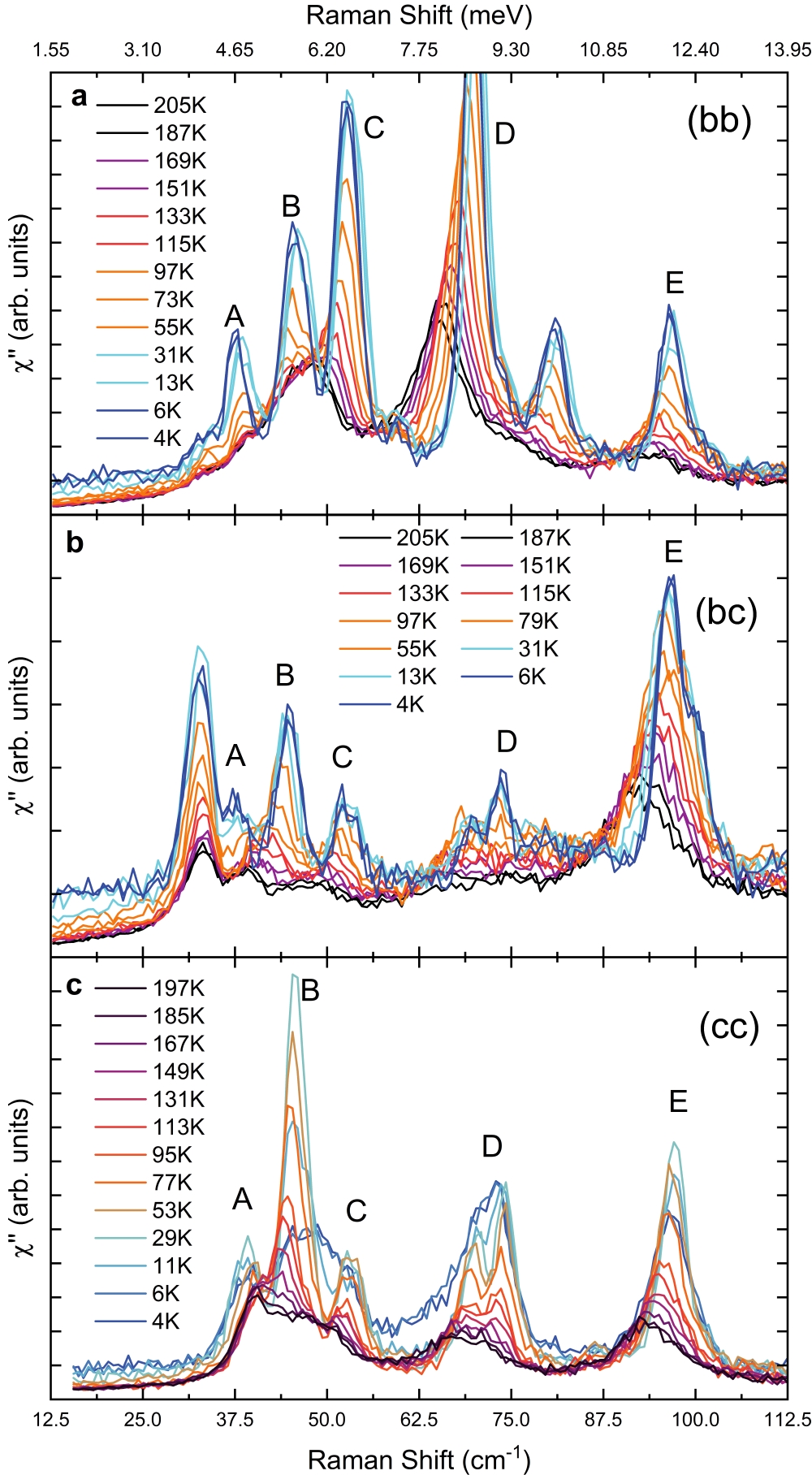}
    \caption{\label{low_energy} Temperature dependence of the Raman scattering spectra of  $\upkappa$-(ET)$_2$Cu$_2$(CN)$_3$ in the spectral range where lattice phonons are observed (below 110~\cm) in the \textbf{(a)} (b,b), \textbf{(b)} (b,c), and \textbf{(c)} (c,c) scattering channels.}
\end{figure}

\begin{figure}
\includegraphics[width=\linewidth]{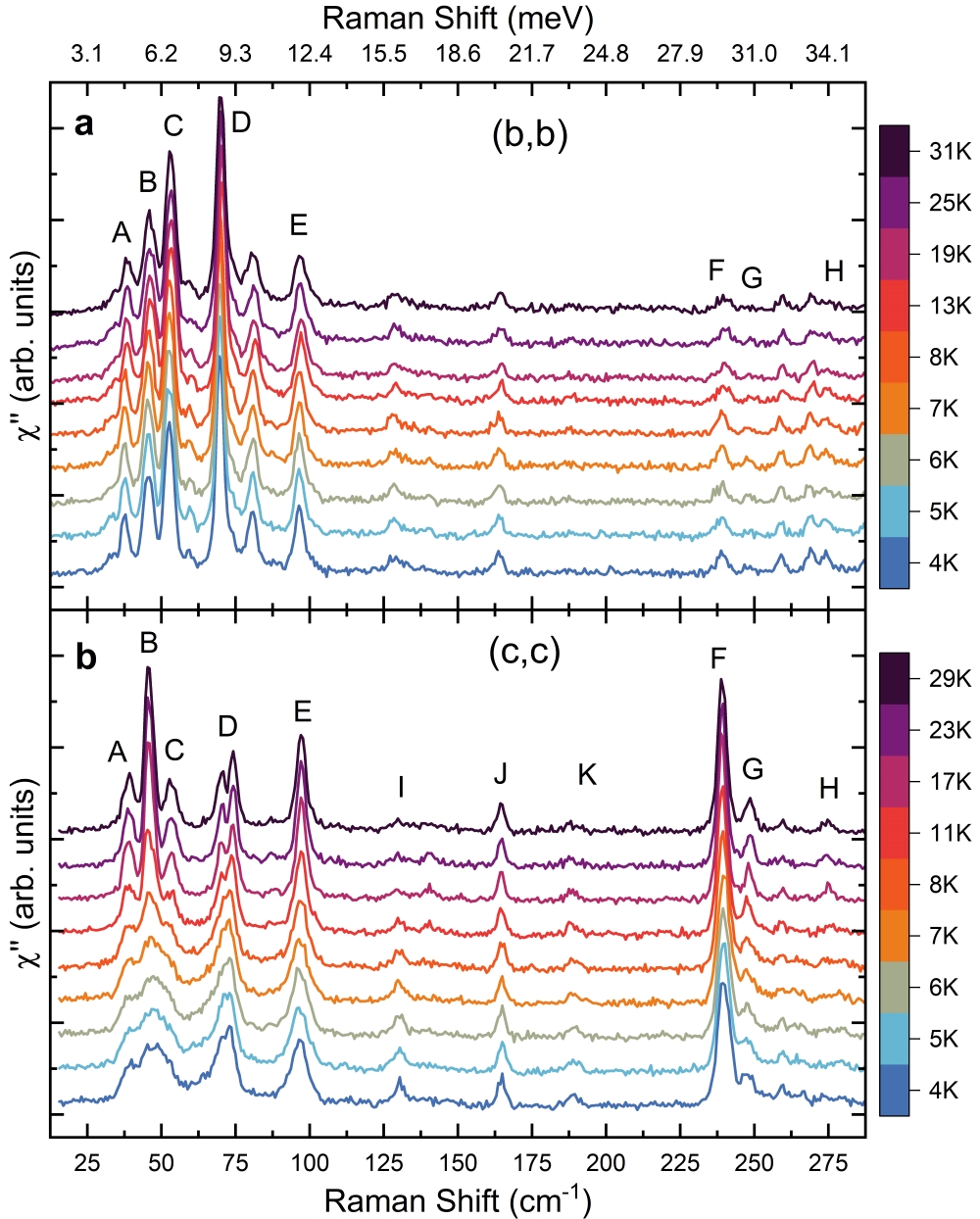}
\caption{\label{Lattice_Low_T} Temperature dependence of the Raman scattering spectra of  $\upkappa$-(ET)$_2$Cu$_2$(CN)$_3$ in the frequency range below 300~\cm\ below 31~K in {\bf (a)} (bb) and {\bf (b)} (cc) scattering channels. Spectra are offset for clarity. Phonons are marked A-H}
\end{figure}

Next, we examine the lattice response of $\upkappa$-Cu-CN by analysing the Raman scattering  spectra in the (b,b), (b,c), and (c,c) scattering channels at low frequencies. Lattice phonons in ET-based compounds are typically observed in the frequency below approximately 100~\cm~\cite{Dressel2016}. In Fig.~\ref{low_energy} we show the Raman spectra in the (b,b), (c,c), and (b,c) polarizations from 12.5~\cm~ to 112.5~\cm, the spectra in the higher frequency range are shown in Fig.~\ref{Lattice_Low_T}, and the assignment of the observed modes based on calculations~\cite{Dressel2016} is presented in SI. One has to keep in mind, that these lattice modes are superimposed on the continuum of the magnetic excitations, which is about an order of magnitude weaker than the observed phonons and detected in low resolution broad range measurements~\cite{Nakamura2014, Hassan2018Crystals}.

\begin{figure}
    \includegraphics[width=\linewidth]{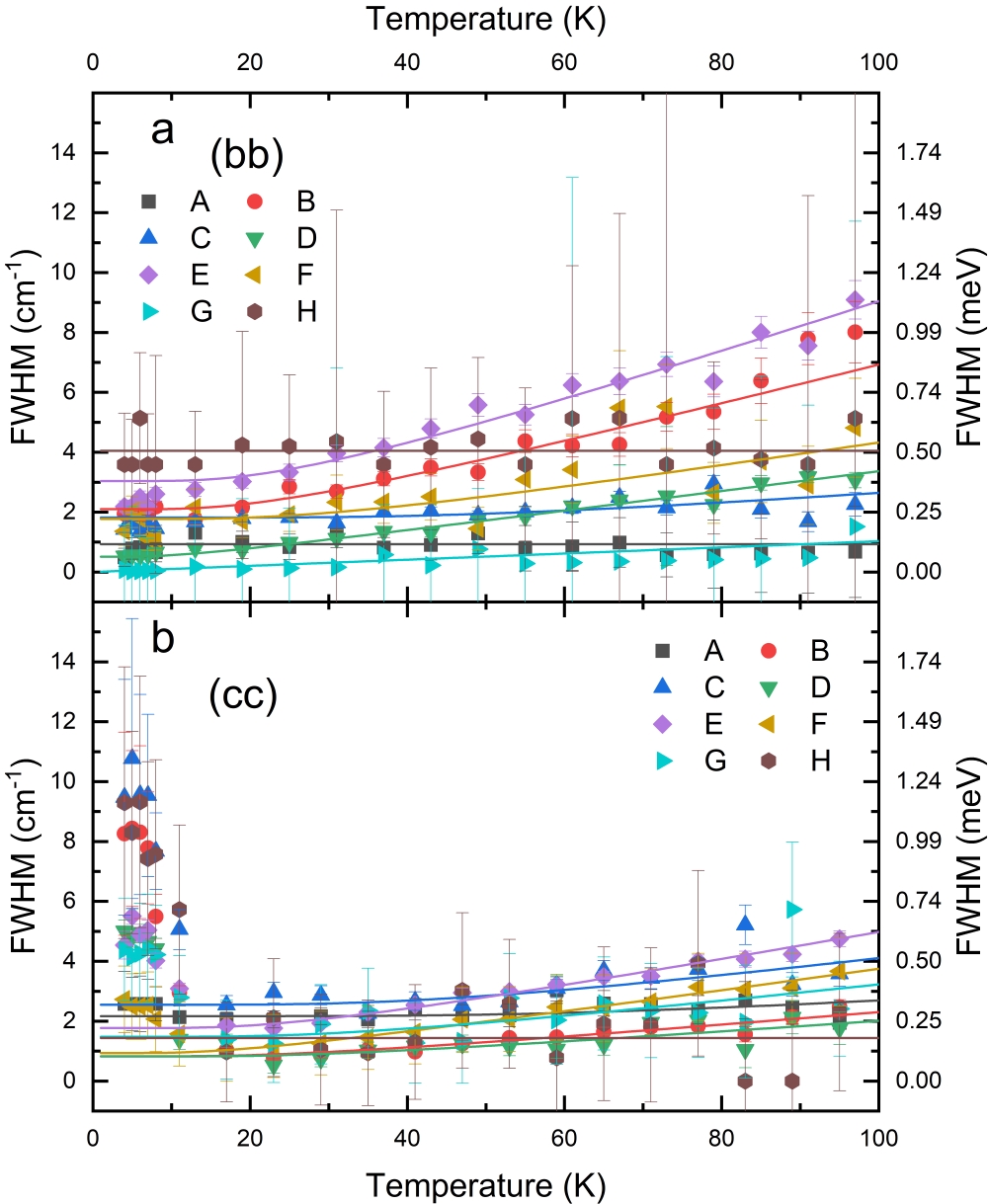}
    \caption{\label{Width} Full-Width-Half-Maximum of modes of lattice vibrations of $\upkappa$-(ET)$_2$Cu$_2$(CN)$_3$ A-H (see notations in Fig.~\ref{Lattice_Low_T}) in \textbf{(a)} (b,b) and \textbf{(b)} (c,c) polarizations. The solid lines are fits according to Ref.~\cite{Klemens1966}, where the fits are performed with data from temperatures above 30~K.}
\end{figure}

The lattice modes in this spectral region are overdamped and weak at room temperature, but can be distinctly observed at temperatures below approximately 150~K.
The changes of phonons on cooling in the (b,b) and (b,c) channels are  conventional: The phonons harden and narrow on cooling, their width described by anharmonic decay according to the Klemens model~\cite{Klemens1966} (see Fig.~\ref{Width}{\color{blue} \bf a}). All the modes increase in intensity on cooling. In (b,c) channel we observe an appearance of  the mode A at 38~\cm~below approximately 20~K.

In the (c,c) scattering channel we observe conventional behavior of the width and frequencies of the lattice modes only down to approximately 20~K.
Below 20~K the modes labeled A-H broaden and lose coherent spectral weight, with the strongest effect observed for the mode B at 46~\cm~(Fig.~\ref{low_energy}c). The linewidth increases on cooling below 20~K to 6~K. (Fig.~\ref{Width}{\color{blue} \bf b})
Several of the modes, notably B, C, and D become so broad and incoherent below 10~K that one cannot isolate their line shape. The fit parameters are therefore not definitive, but represent a best fit with appropriate uncertainties. 

All of the modes A-H correspond to motion of ET molecules~\cite{Dressel2016}. For some there is coupled anion motion, but the mode F corresponds to purely ET motion. 

The anisotropy of the   broadening of phonons at low temperatureswhich is   observed only in the (c,c) polarization, is the most striking result. We do not observe recovery of the width the phonons down to the lowest measured temperature of 6~K, while neutron scattering results, though measured with lower energy resolution, were suggesting such a drop below 6~K~\cite{Matsuura2022}.

\section{Discussion}

The local probe of charge on ET molecules provided by the analysis of line shape of  the molecular vibration $\nu_2$, compared to the  the behavior of the lattice phonons which reflect the collective response, reveals the following picture. Below 40-60~K, the local probe signals an increasing amplitude of charge fluctuations. Both the onset temperature of charge separation of 60~K and its amplitude, estimated at $\Delta n$=0.06$e$, observed by us is in agreement with the dielectric measurements~\cite{Abdel2010,Pinteric2014} and the broadening of the charge sensitive mode $\nu_{27}$ observed by IR measurements~\cite{Sedlmeier2012}.

On further cooling,  we can identify two characteristic temperatures:  At 20~K lattice modes in (c,c) channel demonstrate
broadening, which saturates at 6~K. This coincides with the appearance of mode A at 38~\cm\ in (b,c) channel, demonstrating the loss of local inversion symmetry. These temperature scales also correspond to the ones observed in the measurements of dielectric properties. Below 20~K, a freezing out of the high-temperature dielectric process with glassy relaxation is suggested~\cite{Pinteric2014}, where the 6 ~K anomaly could be a transition into spin-singlet glassy state~\cite{Miksch2021, Riedl2019}.

The anion layer was previously suggested as a source of the disorder potential~\cite{Nakamura2021,Pinteric2014}, which can be a candidate for the origin of a glassy state. However, we cannot attribute the lattice phonons broadening to the structural disorder: The increase in linewidth of the lattice modes is observed only in the (c,c) channel and only in the (ET)$_2$ phonons, without broadening vibrations of the anion layer, such as the lattice modes I, J, and K.
This anisotropy suggests that the origin of the charge order fluctuations and the anisotropic broadening of phonons are competing interactions in this complex material.

\begin{figure}[h]
    \includegraphics[width=\linewidth]{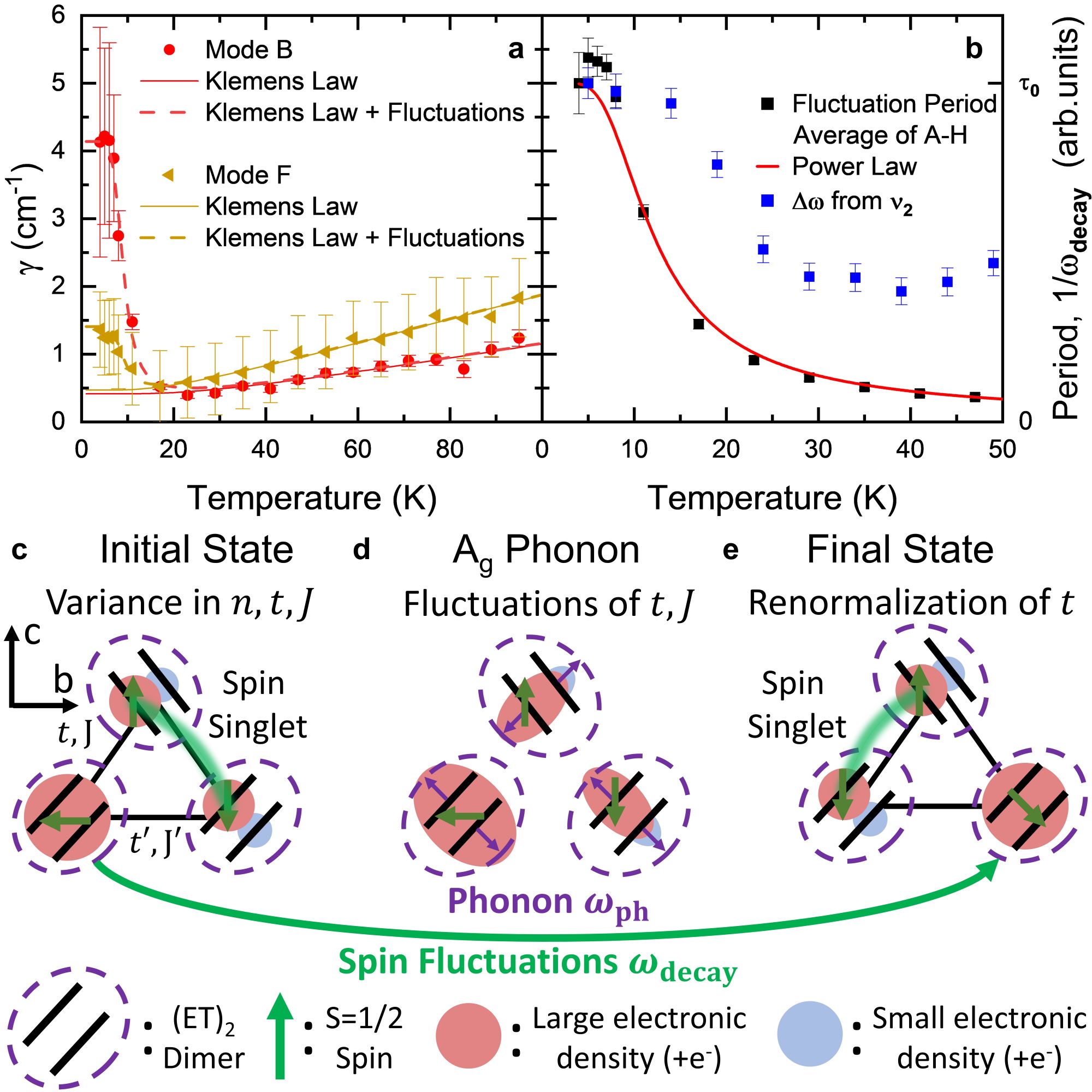}
    \caption{\label{Cartoon} Interaction of fluctuations and phonons. {\bf (a)} The decay rate of modes B and F from data, conventional decay, and additional scattering from fluctuations given fluctuation Lifetime, shown in {\bf (b)}. 
    {\bf (c), (d), (e)} Scattering of defects by phonons. {\bf (c)} An initial state of the valence bond solid is shown. Localized charge on molecular sites (small orange ovals) result in spin singlets (green lines). Dimers which do not have a dipole moment (large orange ovals) represent defects of the dipole glass, with unpaired spins. {\bf (d)} A Raman Active phonon with (cc) Polarizability stretches and compresses the density of states of holes (+e$^-$) along the c axis, corresponding to fluctuations of $t$ and $J$. {\bf (e)} A final state with unpaired spins scattered within the lattice and no change to the net electric dipole. As unpaired spins scatter and dipole moments fluctuate, spin singlets are broken and reformed.}
\end{figure}

Theoretical understanding of charge ordering in $\upkappa$-ET-based dimer Mott insulators has been suggested either within an extended Hubbard model for 1/4 filled bands with electron-phonon coupling~\cite{Dayal2011}  
or within a model which maps the charge degree of freedom within dimer lattice sites onto a S=1 Ising model coupled to S=1/2 Heisenberg spin-interactions in a Kugel-Khomskii model~\cite{Hotta2010, Naka2010, Naka2013}. 
The basic parameters of the extended Hubbard model with each ET$^{+0.5}$ molecule counted as a lattice site  are electron repulsion $U, V$, and charge transfer integrals $t$, where $t_d$ is a charge transfer integral between the molecules in a (ET)$_2$ dimer~(see Fig.~\ref{Cartoon}). The hybridization of the ET HOMO  within (ET)$_2$ dimers renormalizes nearest neighbor repulsion $V_d$ within a dimer to an effective on-site Coulombic interaction:
\begin{equation}\label{Eq 5}
U_{\text{eff}} = 2t_d+V_d+\frac{U-V_d}{2}(1-\sqrt{1+(\frac{4t_d}{U-V_d})^2})
\end{equation}
This allows a mapping of a 1/4-filled model  on  1/2-filling with  the Hamiltonian describing interactions between dimer lattice sites:
\begin{equation}\label{Eq 6}
\begin{split}
H_1 = \ & \ U_{\text{eff}}\sum_{\text{sites }i}  n_{i\uparrow}n_{i\downarrow} + \sum_{i\neq j}V_{ij}(n_{i\uparrow} + n_{i\downarrow})(n_{j\uparrow} + n_{j\downarrow})\\
& - \sum_{i\neq j}\sum_{\text{Spin }\sigma}t_{ij}(c^{\dagger}_{i\sigma}c_{j\sigma} + \text{h.c.})
\end{split}
\end{equation}
where (next) nearest neighboring lattice sites have Coulombic repulsion ($V'$) $V$ and charge transfer integrals ($t'$) $t$. The spin interactions are described by magnetic exchange parameters $J=\frac{4t^2}{U_{\text{eff}}}$ and $J'=\frac{4t'^2}{U_{\text{eff}}}$~\cite{Hotta2010}. 
The Dimer-Mott Insulator to Charge-Order Insulator transition has been parameterized via competition between intramolecular hopping ($t_d$) and interdimer Coulomb Repulsion ($V$)~\cite{Dayal2011, Naka2010, Hotta2010}, and the anisotropy of interdimer hopping ($t'/t$)~\cite{Dayal2011}. In $\upkappa$-Cu-CN, $t'/t$ is slightly above 0.8~\cite{Nakamura2009, Kandpal2009, Jeschke2012}, where numerical simulations cannot predict charge order~\cite{Dayal2011}. 
The most basic tuning of the dipole liquid to dipole solid transition is provided by the increase of  $V/t_d, V'/t_d$ parameters~\cite{Hotta2010, Naka2010}. In the dipole solid phase (static charge order) $t,t'$ and $V,V'$ are renormalized due to unequal occupation within (ET)$_2$ dimers, resulting in electronic anisotropy~\cite{Jacko2020}. 

To the best of our knowledge, theoretical considerations have not accounted for {\it dynamic} renormalization of transfer integrals by charge fluctuations in a dipole liquid state. We expect that this modulation of transfer integrals would result in electron-phonon coupling to the lattice phonons.  Lattice phonons A-H modulate the distance between the planes of ET molecules, the angle between ET molecules, and the relative orientation of ethylene endgroups. Dynamic changes to these structural parameters with modulation of charge transfer integrals~\cite{Mori1984} is the microscopic mechanism of this electron-phonon coupling.
 
 The emerging charge disproportionation  occurs within (ET)$_2$ dimers, so at his point we do not consider fluctuations in the amplitude of $t_d$.
Charge fluctuations dynamically renormalize the charge transfer integral $t$~\cite{Jacko2020}, thus resulting in  corresponding fluctuations of $t$:
\begin{equation}\label{Eq 9}
\langle \delta t (\tau) \delta t (0) \rangle = \Delta_t^2 \exp(-\omega_{\text{decay}}\tau) 
\end{equation}
where $\Delta_t^2 = \langle (\overline{t}-t)^2\rangle$ and $\overline{t}$ is the average of $t$, and  $\omega_{\text{decay}}$ is  the average decay rate of charge fluctuations.  

In a limit where fluctuation frequency is small, $\omega_{\text{decay}} \textless \omega_{\text{ph}}$ (see Fig.~\ref{Mol Vib} c),  we  can express phonon frequency $\omega_{\text{ph}}$ as a function of the change of transfer integral $\delta t$. 
\begin{equation}\label{Eq 7}
\omega_{\text{ph}}(\delta t) = \omega_{\text{ph}}^{(0)} + \delta t \frac{\partial \omega_{\text{ph}}}{\partial t}+ O(\delta t^2)
\end{equation}
where $\omega_{\text{ph}}^{(0)}$ is a lattice phonon frequency at the average value of $t$, or $\delta t = 0$.   $\frac{\delta \omega_{\text{ph}}}{\delta t}$ is a strength of electron-phonon coupling which can be different for every lattice mode. Such coupling will result in $\Delta\omega_i$ a  deviation in frequency arising from the amplitude of charge transfer integral fluctuations $\Delta_t$, so that
$\Delta \omega_i^2 =\langle (\omega_{i}^{(0)}-\omega_{i})^2\rangle = (\frac{\partial \omega_i}{\partial t}\Delta_t)^2$. 
 
The line width of the lattice vibrations can be expressed then as
\begin{equation}\label{Eq 10}
\gamma_i = \gamma_i^{(0)} + C_i^2\frac{\Delta\omega_i^2}{\omega_{\text{decay}}}
\end{equation}
where $\gamma_i^{(0)}$ is the ``natural'' line width unperturbed by the charge fluctuations~\cite{Klemens1966}, $C_i$ is a constant different for every lattice mode.  Lorenzian shape of lattice vibrations A-H in $(c,c)$ scattering channel suggests that the broadening due to the  decreased life-time determined by the  $\omega_{\text{decay}}$ of charge fluctuations is a larger effect for the lattice modes than  $\Delta\omega_i$ at least above 20~K. This is in contrast to the $\nu_2$ mode, which is very sensitive to the redistribution of changes on the ET molecule and changes of $t$ as the result, meaning large $\frac{\delta \omega_{\text{ph}}}{\delta t}$ and a Gaussian line shape. 

Fig.~\ref{Cartoon}a shows the temperature dependence of the line width $\gamma$ of two lattice modes, B and F, in (c,c) scattering channel. At higher temperatures the modes narrow on cooling following the Klemens model, while the broadening below 20~K is determined by the increased scattering on charge fluctuations, as suggested above. 

For each mode, a deviation from the Klemens' Law due to the enhancement of  a phonon scattering rate is 
\begin{equation}\label{Eq gamma'}
 \gamma_i'=\gamma_i-\gamma_i^{(0)}=C_i^2\frac{\Delta\omega_i^2}{\omega_{\text{decay}}}=C_i^2\Delta\omega_i^2\tau_{\text{decay}}
 \end{equation}
  In order to determine the  $\tau_\text{decay}$ independently of $C_i$ and  $\Delta\omega_i$  which we assume to be temperature independent for the modes A-H,   we can express 
  $$\tau_\text{decay}(T)=\frac{1}{N}\sum_{i=1}^N\frac{\gamma_i'(T)}{\gamma_i'(6~\text{K})}$$
  Such averaging provides a reliable result for the temperature dependence of $\tau_{\text{decay}}$.  Fig.~\ref{Cartoon}b demonstrated that the fluctuation life time increases as temperature decreases and reaches saturation as  temperatures approach 6~K.  

We have to keep in mind, that if fluctuation frequency is small enough $\omega_{\text{decay}} \ll \omega_{\text{phonons}}$ we would not be able to distinguish it from a fully static case. A new phonon at 38~\cm continuously increases the intensity  in $(b,c)$ scattering channel below 20~K, suggesting possible local symmetry breaking.

The extreme broadening of low frequency modes as temperature approaches 6 K (see Fig.~\ref{Lattice_Low_T}) suggests the higher electron-phonon coupling $\frac{\delta\omega_i}{\delta t}$ for these modes and an importance of this term at low frequencies. The increase of $\Delta t$ is in agreement with the increase of $\Delta n$ the amplitude of charge disproportionation probed by the $\nu_2$ mode below approximately 20~K (see Fig.~\ref{Mol Vib}). 

The coupling of phonons to the charge fluctuations as discussed above is observed only in (c,c) scattering channel. Such anisotropy suggest anisotropic charge fluctuations. The anisotropy detected by us is in agreement with the stronger dielectric response in E$\parallel$c ~\cite{Pinteric2014} and enhanced optical conductivity, where an additional electronic response, strongly coupled to lattice phonons, increases on cooling below approximately 20~K in E$\parallel$c polarization~\cite{Dressel2016,Itoh2013}. While the authors of Ref.~\cite{Dressel2016} have left the interpretation of the low frequency continuum open, the best candidate for these excitations are the charge order fluctuations. 

Previously, strong anisotropy of the electronic structure was observed only in a charge ordered state of a  $\kappa$-phase material by infrared and Raman  spectroscopies ~\cite{Drichko2014,Hassan2020} and was in agreement with the theoretically demonstrated renormalization of overlap integrals in the charge ordered state ~\cite{Jacko2020}. Our results as well as Ref.~\cite{Dressel2016,Itoh2013,Pinteric2014} show that already in a fluctuating regime charge disproportionation shows some anisotropy. 

\begin{figure*}
    \centering
    \includegraphics[width=\textwidth]{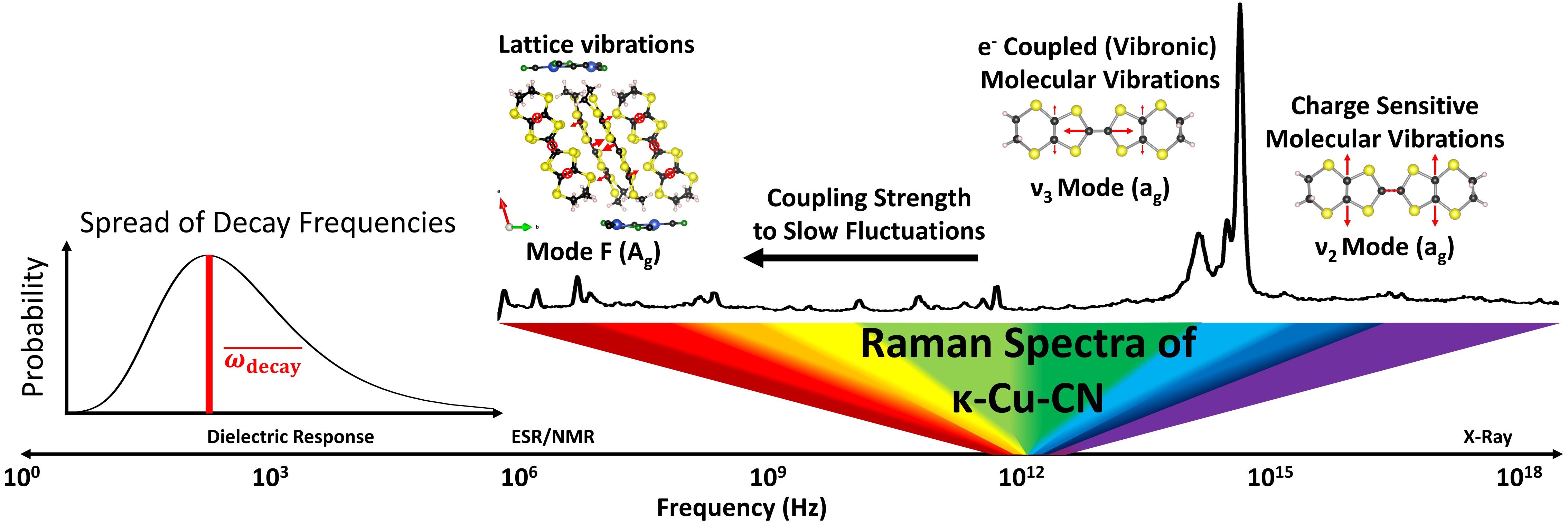}
    \caption{A scheme summarizing the frequency range of methods used to probe electrodynamic and magnetic response of $\upkappa$-(ET)$_2$Cu$_2$(CN)$_3$.  Charge fluctuations have a broad distribution of frequencies $\omega_{\text{decay}}$, with higher part of distribution possible reaching  the frequencies of lattice vibrations. The decay process couples more weakly to molecular vibrations, which have more than order of magnitude higher frequencies than lattice phonons.}
    \label{Freq. Fig}
\end{figure*}

Our conclusions are in agreement with the dielectric measurements in  Ref.~\cite{Pinteric2014}, which suggest a  slow down of dipole fluctuations on cooling the samples from 60~K. Mean relaxation time of the dielectric response 10$^{-8}$~s observed at high temperatures decreases down to the average timescale of 10$^{-5}$~s. While the relaxation time at around 50~K is close to the characteristic time of lattice phonons, 10$^{-11}$~s, the low-temperature behavior where  $\omega_{\text{decay}} \ll \omega_{\text{ph}}$ indeed could be indistinguishable from static regime by Raman scattering. In comparison of the results obtained in different frequency ranges (Fig.~\ref{Freq. Fig}), we have to keep in mind that in any relaxation process described by correlation functions is stochastic. Our correlation function, which is determined by an exponential decay, has an average timescale of $\tau_{\text{decay}}$ and standard deviation of decay times $\sigma_\tau=\sqrt{\langle (\tau-\tau_{\text{decay}})^2\rangle} = \tau_{\text{decay}}$\cite{Kingman1992}. The increase of $\tau_{\text{decay}}$ naturally leads to the increase of $\sigma_\tau$, which can cover a range of frequencies from dielectric to Raman and optical measurements, see Fig.~\ref{Freq. Fig}. Additionally, a large distribution of frequencies is a natural consequence of localized defects emerging from disorder. Such broadening of the characteristic time scales as well as a broad frequency range of the spectroscopic probes applied to the studies of $\upkappa$-Cu-CN (see Fig.~\ref{Freq. Fig}  ) can explain the rich and at times  conflicting experimental results for this material. At finite temperatures below 6~K, defects corresponding to structure, dipoles, and spin can exhibit responses that cover a wide range of frequencies, resembling gapless features.

Extended Hubbard Model is a framework to understand a transition from Dimer Mott Insulator to Charge order, with fluctuations of electrical dipoles and intrinsic inhomogeniety of a system arising from competing interactions close to a phase transitions~\cite{Yamashita2020,Urai2022}. Our experimental data suggest that a glassy state formed by the  dipole moments within the (ET)$_2$ dimers can be additionally a consequence of the electron-phonon coupling in $\upkappa$-Cu-CN. For understanding of charge and magnetic state in this material, a model, for example a Kugel-Khomskii model \cite{Kugel1982}, used to describe this transition, should include dynamic electron-phonon coupling.

Some models which describe similar physical effects already exists. With electric dipoles on (ET)$_2$ dimers being mapped on an Ising model {\cite{Hotta2010, Naka2010, Naka2013}}, it is useful to relate to recent theoretical work on disordered Transverse Ising Model.  In our case, the disorder can be present both in the amplitude and orientation of Ising-like dipoles.  It was shown that when bond randomness is introduced to the Transverse Ising model, it can stabilize a structural-spin-glass state~\cite{Hotta2022, Hotta2023}. This  is consistent with our experimental results of disordered electric dipoles  with strong coupling to the lattice in $\upkappa$-Cu-CN. The model has been already successfully applied to different experimental systms, the pyrochlore molybdates, an Ising systems where the Jahn-Teller effect is expected to drive glassy dynamics~\cite{Mitsumoto2022}. These experimental systems carry  similarity to $\upkappa$-Cu-CN, for example in cooling rate effects~\cite{Yakushi2015}.  Such so called ``quantum-mechanically driven structural-spin-glass" exhibits non-trivial thermodynamics, such as a Berezinskii-Kosterlitz-Thouless transition~\cite{Hotta2022}, which can be further studied in $\upkappa$-Cu-CN.

It is important to note, that the observed by us anisotropic broadening of lattice phonons is so far unique for $\upkappa$-Cu-CN among molecular Mott insulators despite charge fluctuations observed in other systems~\cite{Hassan2018,Urai2022,Hassan2018Crystals}. This could be a consequence of the difference in electronic parameters and anisotropy, which also results in the smallest observed $\Delta n$ charge disproportionation in $\upkappa$-Cu-CN. Magnetic exchange Raman scattering observed  in the spectra of $\upkappa$-Cu-CN ~\cite{Nakamura2014,Hassan2018,Hassan2018Crystals} shows that despite finite $\Delta n$, the system still behaves as S=1/2 Mott insulator (Dimer Mott insulator) on a triangular lattice, in contrast to other organic Mott insulators with charge fluctuations and charge order~\cite{Hassan2018,Yamashita2020,Drichko2022}.

Since the magnetic exchange is $J = \frac{4t^2}{U_{\text{eff}}}$, we expect that fluctuations of $t$, $\Delta_t$ correspond to fluctuations of $J$, $\Delta_J = \frac{8t}{U_{\text{eff}}}\Delta_t$.  These fluctuations, which have a broad range of low frequencies   below 6~K, can explain the origin of the disordered magnetic ground state and spin singlet pairs with life time of $\tau_{\text{decay}}$, as measured in ESR~\cite{Miksch2021}, NMR, and magnetic torque~\cite{Isono2016,Riedl2019}, and remind the RVB model. 

In summary, we used Raman spectroscopy to probe the electronic state and lattice dynamics of $\upkappa$-Cu-CN. We observe a developing  of slowly fluctuating charge disproportionation  at temperatures below 60~K. On temperature decrease, the charge disproportionation increases in amplitude and the fluctuations slow down. We observe strong and highly anisotropic coupling of these charge   fluctuations to the lattice. Our observation are in agreement with a large number of measurements of electrodynamic response of $\upkappa$-Cu-CN, which probed some manifestations of this complex effect. We review theoretical work which shows that this coupling of  dipole fluctuations to the lattice can lead to a formation of glassy state. Since dipole fluctuations dynamically renormalize magnetic interactions, our observation suggest the understanding of magnetic properties of $\upkappa$-Cu-CN as slowly fluctuating spin dimers, reminiscent of RVB state.

\section{Acknowledgments}
The authors are grateful to C. Hotta and S. Winter for fruitful discussions. The work at JHU was supported by NSF Award No. DMR-2004074. This work was performed in part at Aspen Center for Physics, which is supported by NSF grant PHY-2210452.
 
\bibliography{./main.bbl}

\end{document}


\title{Supplementary Information for Novel Dipole-Lattice coupling in the Quantum-Spin-Liquid Material $\upkappa$-(BEDT-TTF)$_2$Cu$_2$(CN)$_3$}

\author{Jesse Liebman}

\affiliation{The Department of Physics and Astronomy, The Johns Hopkins University, Baltimore, Maryland 21218, USA}
\author{Kazuya Miyagawa}

\affiliation{University of Tokyo, Hongo, Tokyo, Japan}
\author{Kazushi Kanoda}

\affiliation{University of Tokyo, Hongo, Tokyo, Japan}

\author{Natalia Drichko}
\affiliation{The Department of Physics and Astronomy, The Johns Hopkins University, Baltimore, Maryland 21218, USA}

\date{\today}

\begin{abstract}

\end{abstract}

\pacs{}
\maketitle

\section{Methods}
Raman scattering spectra in the frequency range from 12 to 2000~\cm\ and temperature range from 300 down to 6 K were measured using the Jobin-Yvon T64000 Triple Monochromator Spectrometer equipped with a liquid Nitrogen cooled CCD. Measurements were performed in a psuedo-Brewster angle geometry.  Line of Ar$^+$-Kr$^+$ laser at 514.5nm was used for excitation. The laser power was kept at 1~mW with an elliptical laser probe size of $50 \times 100 \ \upmu$m, to ensure laser heating of less than 2K. For low temperature measurement, samples were glued on a cold finger of  a Janis ST500 Cryostat and cooled from 300~K to 6~K (4 K measured on the cold finger) using a cooling rate of 0.1~K/min. Samples were glued to the cold finger using GE Varnish. To minimize strain caused by the mismatch of thermal contraction of the sample and Cu sample holder,  samples were glued by one  edge, to allow thermal contaction of the sample be unaffected. Measurements were done on 2 samples to ensure reproducibility of results. 
The crystals were oriented to the polarizations of the incident and scattered light using by polarization dependent Raman measurements. Our notation for symmetry group of measured Raman excitations   refers to the structure and symmetry of the BEDT-TTF layer, as opposed to the structure of the whole crystal, which raises the point group symmetry from C$_{\text{2h}}$ to D$_{\text{2h}}$. Raman active molecular vibrations are in the a$_{\text{g}}$ representation in an individual ET molecule (C$_2$). By placing ET molecules into the $\upkappa$-ET$_2X$ lattice (D$_{\text{2h}}$), these modes exhibit a variety of representations (A$_{\text{g}}$, B$_{\text{1g}}$, B$_{\text{2g}}$, and B$_{\text{3g}}$), depending on the relative phase of vibration between ET molecules. We measured with light incident along the $a$ axis of the material, allowing us to probe modes in the A$_{\text{g}}$ and B$_{\text{3g}}$ representations. We assign A$_{\text{g}}$ symmetry to the (b,b) and (c,c) scattering channels and B$_{3\text{g}}$ symmetry to the (c,b) and (b,c) scattering channels.

\section{Fitting Procedure}
We fit features of molecular vibrations and lattice phonons using Voigt profiles, or a convolution of a Gaussian distribution with a Lorentzian profile: 
$$I(\omega) \propto \int_{-\infty}^{\infty}d\omega'G(\omega', \omega, \sigma_\text{res})L(\omega, \omega_i, \Gamma_i)$$ 
The $\sigma_\text{res}$ parameter is the Gaussian spectral resolution parameter. For all measurements, this value is  2~\cm. $\omega_i$ is the frequency of vibration. $\Gamma_i$ is the phenomenological damping factor. 
We fit each molecular vibration and lattice phonon individually and then fit all of the lineshape parameters over a frequency range of 200~\cm ~together with a linear background to approximate broad features, such as photoluminescence and the spinon continuum~\cite{Hassan2018Crystals}.

For the charge sensitive molecular vibration $\nu_2$ we use three different models of line shapes to fit experimental data: a two-site model with  fluctuations between the cites of frequency $\omega_{ex}$, a two-site static model, and a two-site model with Gaussian fluctuations. 

For the two-site model with  fluctuations we use line shape description provided in Ref.~ \cite{Kubo1969, Yakushi2012_Crystals}

\begin{widetext}
\mbox{$I(\omega) \propto \text{Re} \left [ \begin{pmatrix} a_1 & a_2
\end{pmatrix} 
\begin{pmatrix}
    \mathrm{i}(\omega - (\omega_0 - \frac{\Delta}{2}
    )) + \frac{\omega_{\text{ex}}}{2} + \Gamma_0/2 & -\frac{\omega_{\text{ex}}}{2}
    \\
    -\frac{\omega_{\text{ex}}}{2} & \mathrm{i}(\omega - (\omega_0 + \frac{\Delta}{2}
    )) + \frac{\omega_{\text{ex}}}{2} + \Gamma_0/2
\end{pmatrix}^{-1}
\begin{pmatrix}
    a_1 \\
    a_2
\end{pmatrix}
\right ]$}
\end{widetext}

Where $\mathrm{i}$ is the imaginary unit. $a_1$ and $a_2$ are intensity prefactors for sites 1 and 2, respectively. $\omega_0$ is the frequency of $\nu_2$ where charge is equally shared between sites 1 and 2, and $\Gamma_0$ is the phenomenological damping constant of $\nu_2$. $\Delta$ is the splitting in frequency for each site and $\omega_{\text{ex}}$ is the frequency of fluctuations of charge between sites. For the limiting case without fluctuations (static charge order) we employ a sum of two Voigt profiles related by a splitting in frequency, $\Delta$. 

The model with Gaussian fluctuations is approximated by a Voigt profile with the Gaussian standard deviation, $\sigma$, such that $\sigma^2$ = $\sigma_C^2+\sigma_{\text{res}}^2$. $\sigma_C$ represents the magnitude of random fluctuations between sites and their associated decay rate~\cite{Kubo1969, Yakushi2012_Crystals}, and $\sigma_{\text{res}}$ is the resolution parameter.

\begin{figure}
    \centering
    \includegraphics[width=\linewidth]{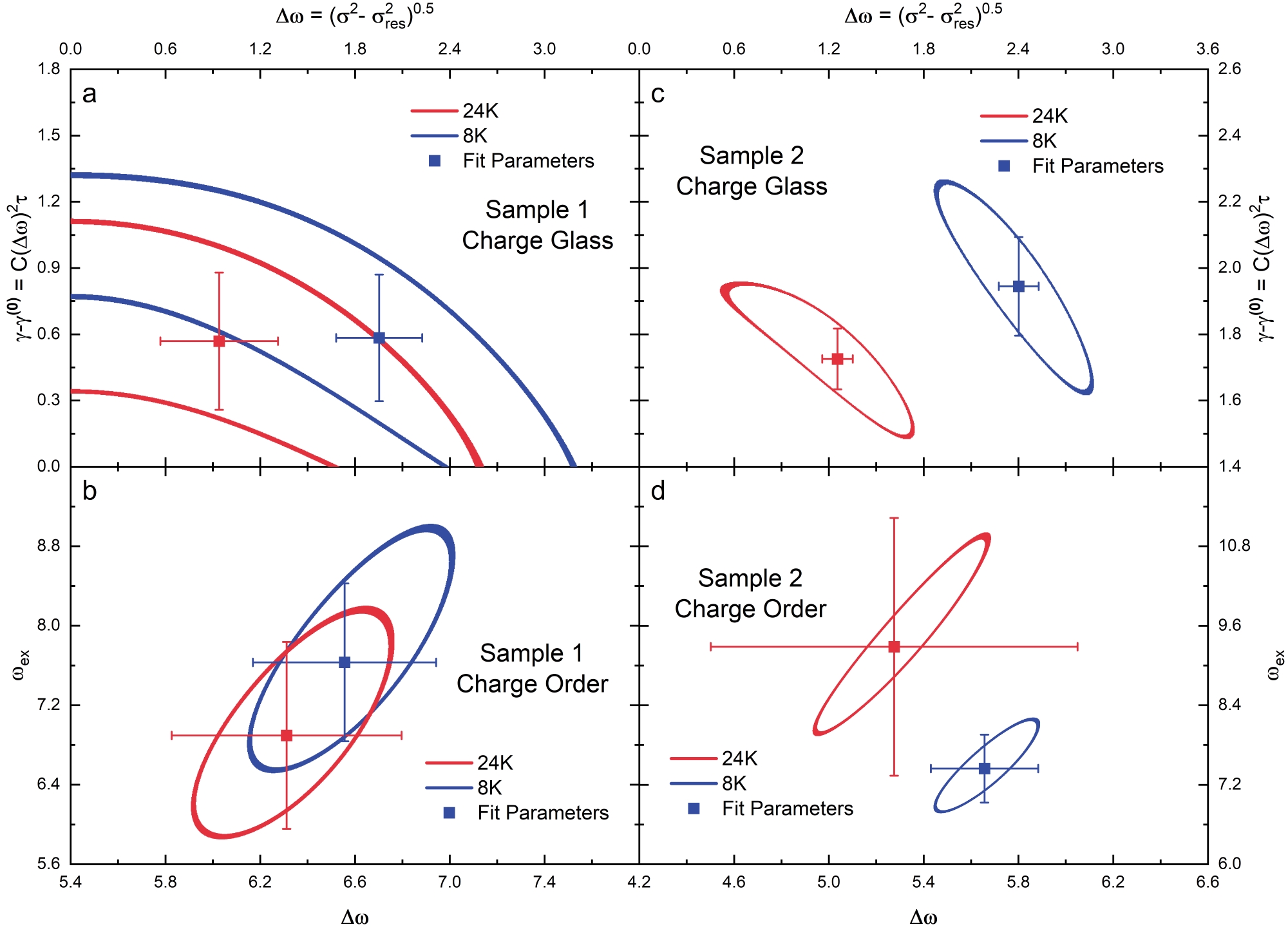}
    \caption{Comparison of confidence regions to estimated error of fit parameters. For Sample 1: {\bf(a)} Charge Glass and {\bf(b)} Charge Order. For Sample 2: {\bf(c)} Charge Glass and {\bf(d)} Charge Order}
    \label{SI_Fig. 1}
\end{figure}

Calculation of confidence regions is estimated using the studentized bootstrap~\cite{Efron1994}. We use the total squared error (TSE) as the estimand, where we compare TSE$_m(x_i)$ for the resampled model parameters $x_i$ to TSE$_m(\hat{x_i})$ for the estimated model parameters. We define the standard error of the TSE for each model $m$ as:\\
$$\hat{\sigma_m}=\sqrt{\frac{1}{N-N_m-1}\sum_{i=1}^N(X_{\text{obs}}-X_\text{m})^2}=\sqrt{\frac{1}{N-N_m-1}\text{TSE}_m(\hat{x_i})}$$
where $X_\text{obs}$ is the observed spectra, $X_m$ is the calculated spectra from model $m$, $N$ is the number of data points, and $N_m$ is the number of parameters of model $m$. We then calculate confidence regions using the following t-score:
$$t(x_i)=\frac{\text{TSE}_m(x_i)-\text{TSE}_m(\hat{x_i})}{\hat{\sigma_m}}$$
We demonstrate the validity of the statistic by comparing estimated error of the fit parameters to the calculated confidence regions. The calculated confidence regions account for the non-linear behavior of the models used to parameterize the spectra. The comparison is shown in~\ref{SI_Fig. 1}
\clearpage
\pagebreak
\begin{table*}[h]
    \scriptsize
    \centering
    \begin{tabular}{c|c|c|c|c|c}
         \hline
         Mode & Mode  & 4K Freq.  & DFT Freq.  & ET Layer Motion & Cu-CN Layer Motion \\ This Work & Dressel et. al.
         & (cm$^{-1}$) & (cm$^{-1}$) & & \\
         \hline
         * & 5 & 32.9(2) & 33.3 & (ET)$_2$ dimers twisting in-phase, & c-axis shear movement\\
          & & & & Breathing in counterphase within (ET)$_2$ & \\
          \hline
         A & 9 & 37.64(7) & 38.2 & Bending in-phase motion of (ET)$_2$ & b-axis shear motion\\
         B & 14 & 45.57(4) & 46.0 & Shear motion & Shear motion\\
         C & 18 & 52.70(2) & 51.7 & Wagging of ET molecules & b-axis motion of anions\\
         D & 33 & 72.7(1) & 72.2 & Wagging of ethylene endgroups & Wagging of Cu bonds; bending of CN links\\
         E & 48 & 97.1(2) & 96.2 & Rocking of ethylene groups & Rotation of Cu bonds; twisting of CN bonds\\
         I & 64 & 130.9(2) & 129.0 & Coupled ET motion & Bending and stretching of anions\\
         J & 76 & 165.3(1) & 165.1 & Coupled ET motion & Bending and stretching of CN\\
         K & 84 & 189.4(3) & 194.9 & Coupled ethylene group motion & Twisting and stretching of anion layer\\
         F & 91 & 240.2(3) & 242.0 & Bending of ET molecules & \textbf{No anion motion}\\
         G & 97 & 248.6(4) & 254.7 & Wagging of ET molecules & Coupled out of plane motion of anions\\
         H & 110 & 275(1) & 273.8 & Rocking of ethylene groups & Coupled anion motion\\
         M & 198 & 638.8(8) & 638.7 & Rocking of ethylene groups & \textbf{No anion motion}\\
         N & 202 & 644.4(7) & 641.8 & Rocking of ethylene groups & \textbf{No anion motion}\\
         \hline
    \end{tabular}
    \caption{Lattice Vibrations and Low Energy Molecular Vibrations in $\upkappa$-(BEDT-TTF)$_2$Cu$_2$(CN)$_3$. DFT frequencies are used for comparison, with listed motion from Ref~\cite{Dressel2016}}
    \label{Table I}
\end{table*}
\begin{figure}
        \centering
        \includegraphics[width=\linewidth]{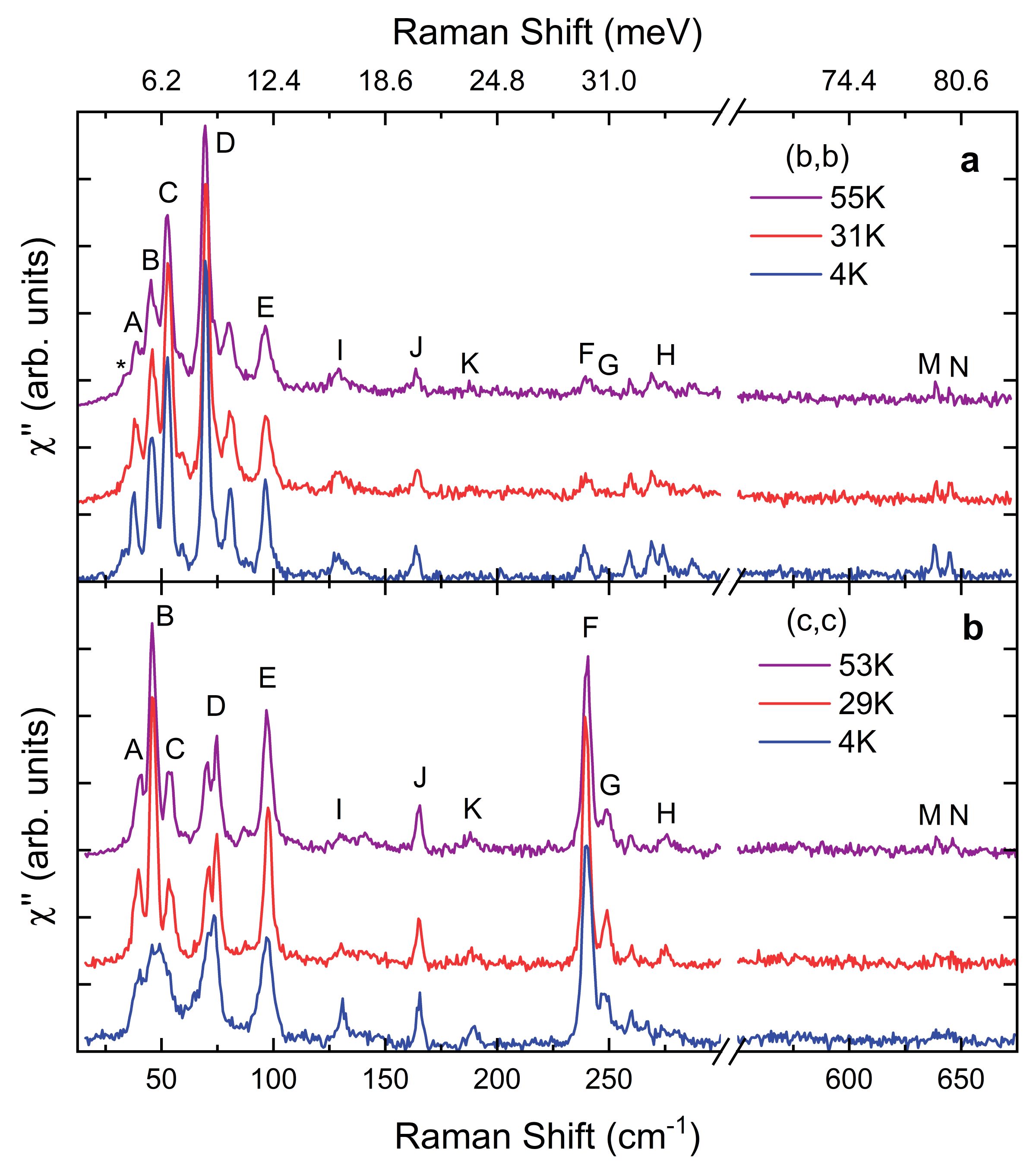}
        \caption{Spectra of Lattice Phonons and Low Energy Molecular Vibrations of $\upkappa$-(BEDT-TTF)$_2$Cu$_2$(CN)$_3$ in the {\bf (a)} (b,b) and {\bf (b)} (c,c) polarizations. Modes are labeled according to Table~\ref{Table I}}
        \label{SI_Triple}
    \end{figure}
\clearpage
\pagebreak
\begin{figure}
        \centering
        \includegraphics[width=\linewidth]{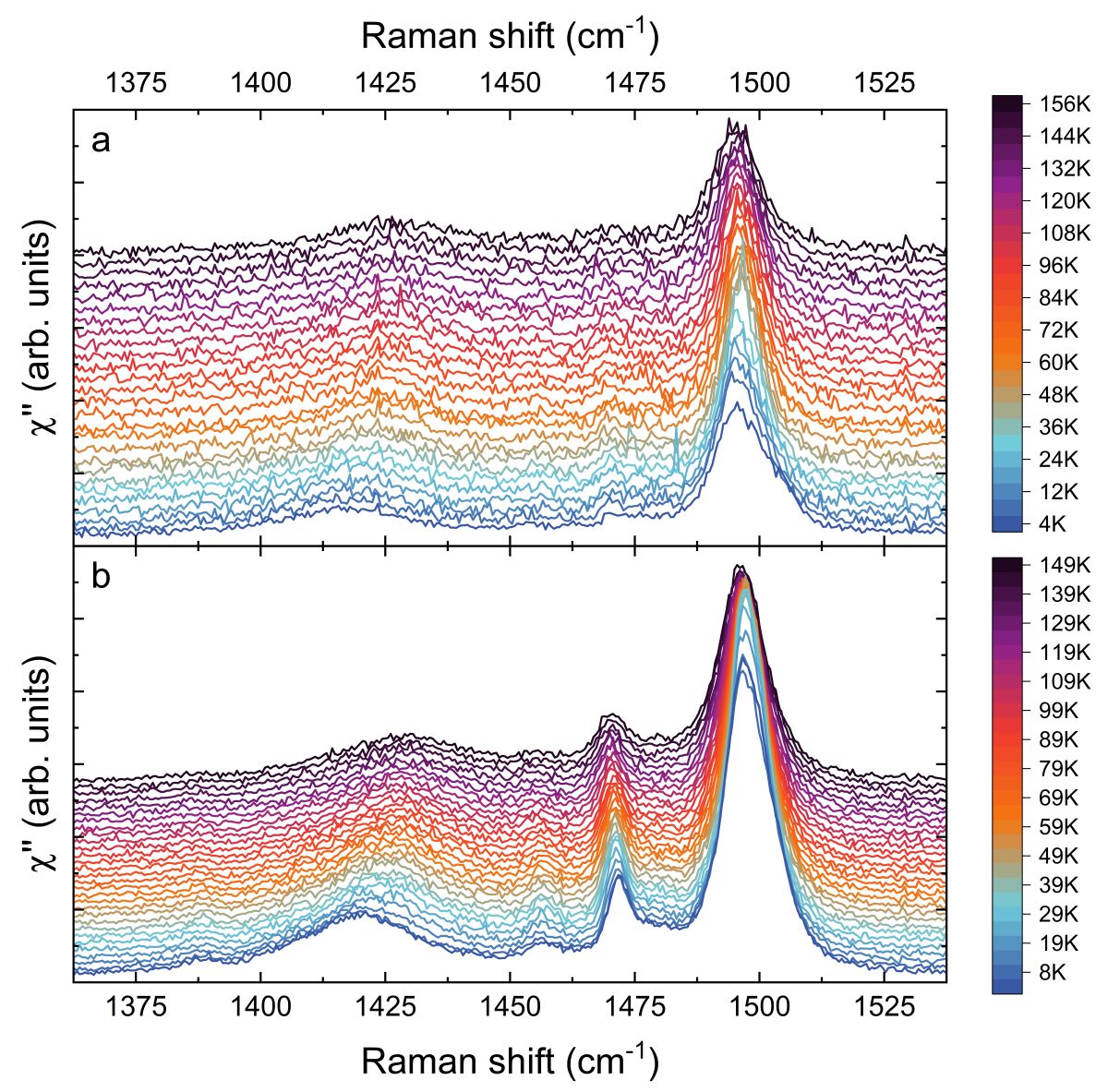}
        \caption{Spectra of Molecular Vibrations of $\upkappa$-(BEDT-TTF)$_2$Cu$_2$(CN)$_3$ in the (bc) polarization for {\bf (a)} Sample 1 and {\bf (b)} Sample 2}
        \label{SI_Single}
    \end{figure}
\clearpage
\pagebreak
\begin{figure}
        \centering
        \includegraphics[width=\linewidth]{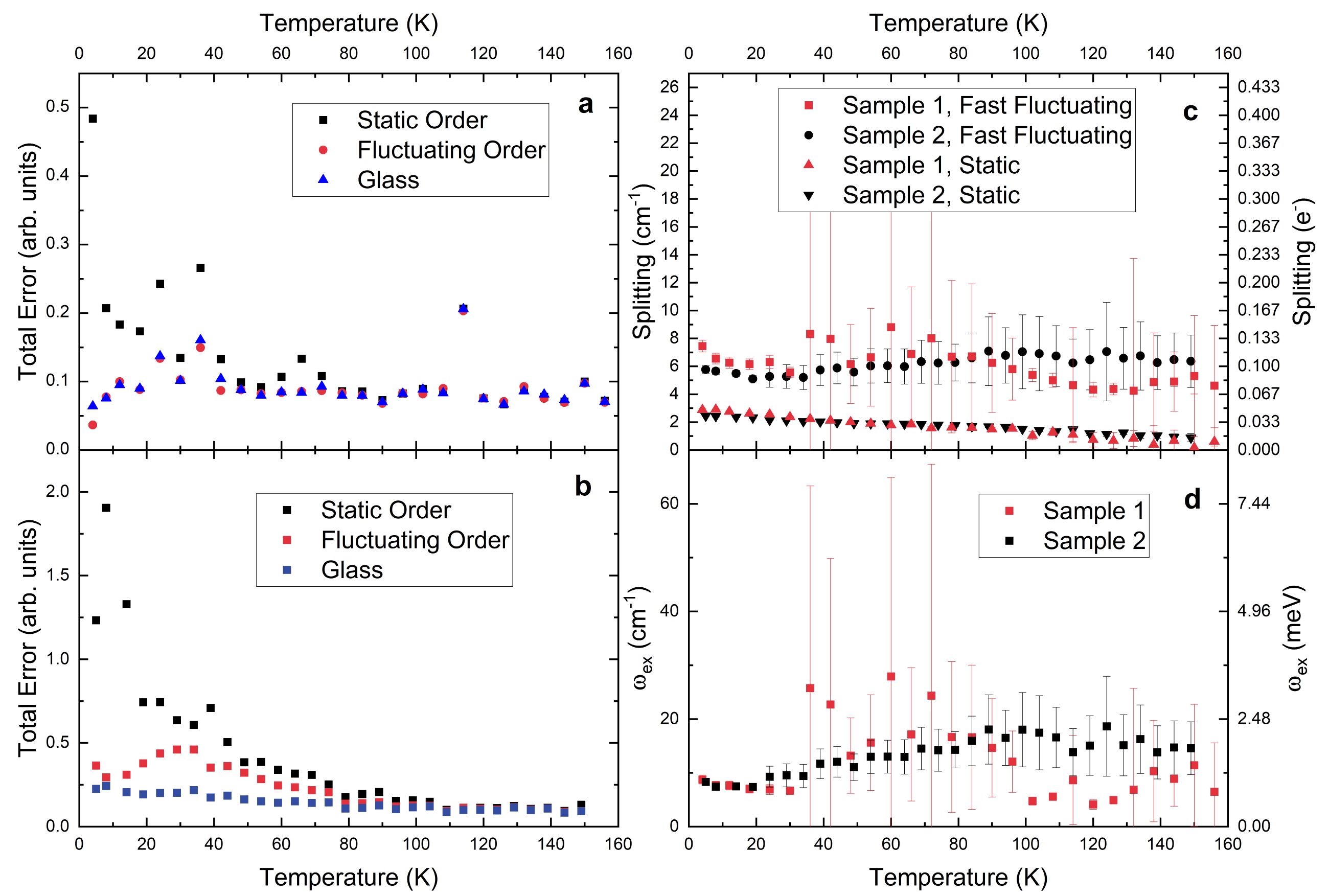}
        \caption{Total Squared Error by Model in {\bf (a)} Sample 1 and {\bf (b)} Sample 2. {\bf (c)} Charge Splitting for Static and Fluctuating Order. {\bf (d)} Exchange Frequency $\omega_{\text{ex}}$ in the two-site model.}
        \label{SI_Mol_Vib}
    \end{figure}
When comparing the models for the $\nu_2$ lineshape, we considered the statstic {\it Total Squared Error} (TSE) as well as the temperature dependence of best-fit parameters. As shown in Fig.~\ref{SI_Mol_Vib} the glass model provides a better fit than the ordered models. Even more importantly is the temperature dependence. Note that for the best fit parameters for the two-site jump model the exchange frequency increases at lower temperatures (Fig.~\ref{SI_Mol_Vib}d). This behavior is at odds with the understanding of the charge degree of freedom. When forcing the model to consider static charges, the fits perform far worse and produce smaller charge disproportionation. The TSE for the ordered model performs worse as $\omega_{\text{ex}}$ decreases. We can therefore reject the ordered model for two reasons: fit-quality and implied charge dynamics.
\clearpage
\pagebreak
\begin{figure}
    \centering
        \includegraphics[width=\linewidth]{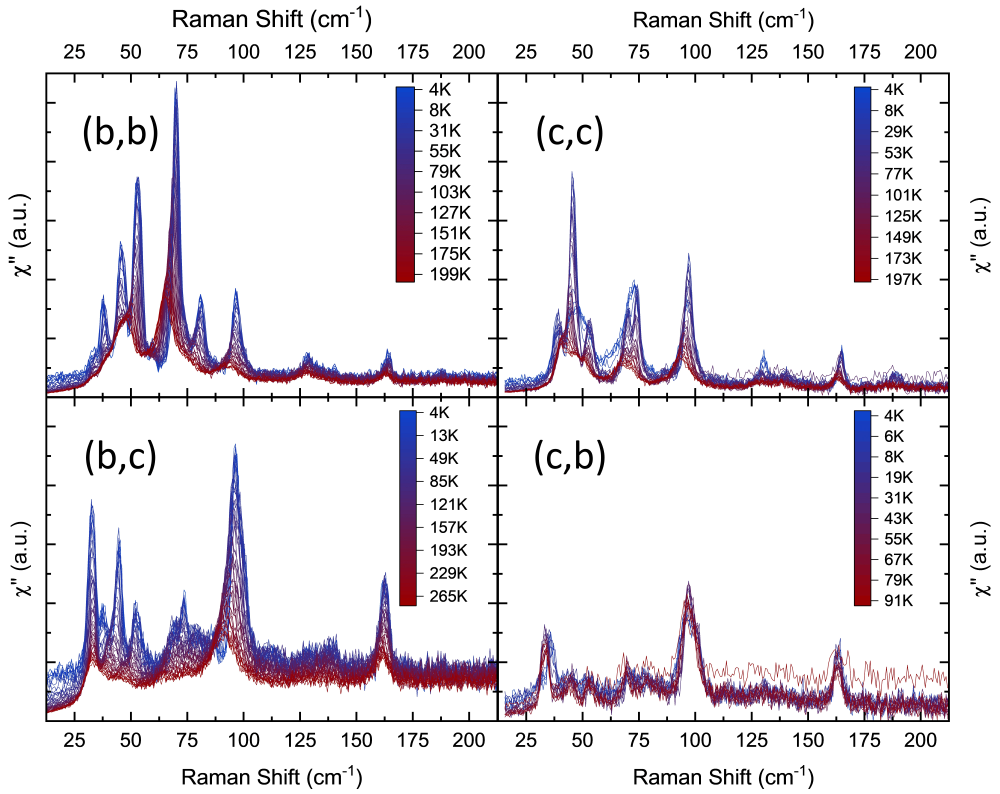}
        \caption{Spectra of low energy lattice phonons of $\upkappa$-(BEDT-TTF)$_2$Cu$_2$(CN)$_3$ in Sample 2 below 200~\cm~ in the {\bf (b,b)}, {\bf (b,c)}, {\bf (c,c)}, and {\bf (c,b)} polarizations.}
        \label{SI_TL2}
\end{figure}
\clearpage
\pagebreak
\begin{figure}
    \centering
        \includegraphics[width=\linewidth]{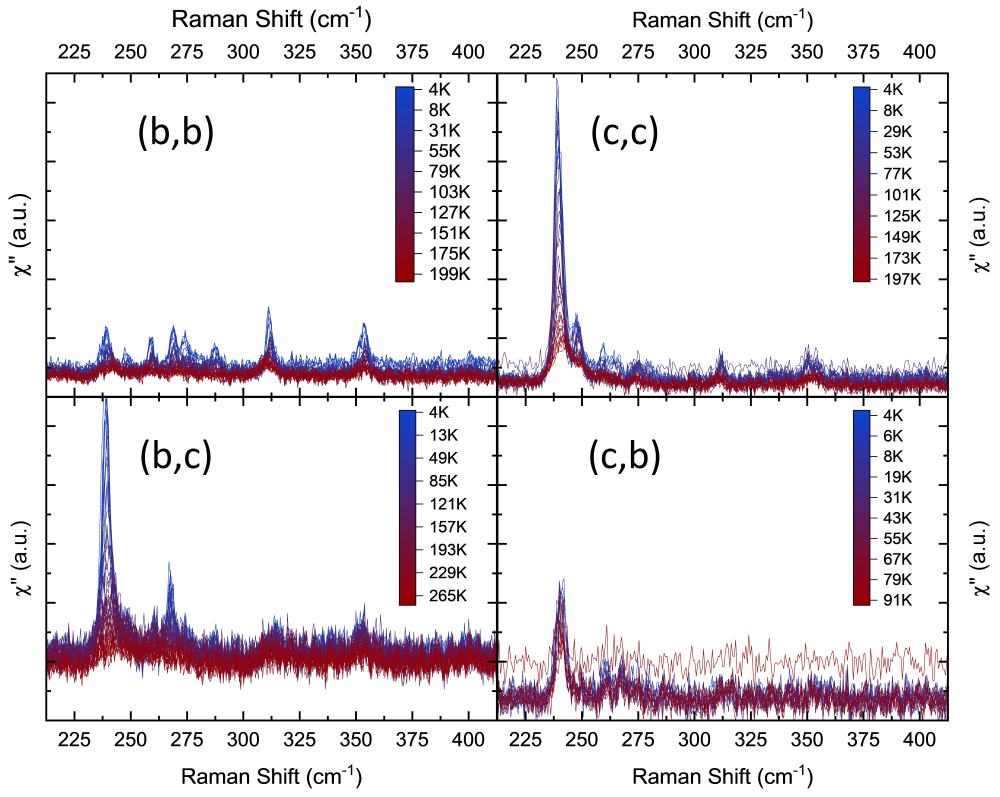}
        \caption{Spectra of lattice phonons of $\upkappa$-(BEDT-TTF)$_2$Cu$_2$(CN)$_3$ in Sample 2 from 200~\cm~ to 400~\cm~ in the {\bf (b,b)}, {\bf (b,c)}, {\bf (c,c)}, and {\bf (c,b)} polarizations.}
        \label{SI_TM2}
\end{figure}
\clearpage
\pagebreak
\begin{figure}
    \centering
        \includegraphics[width=\linewidth]{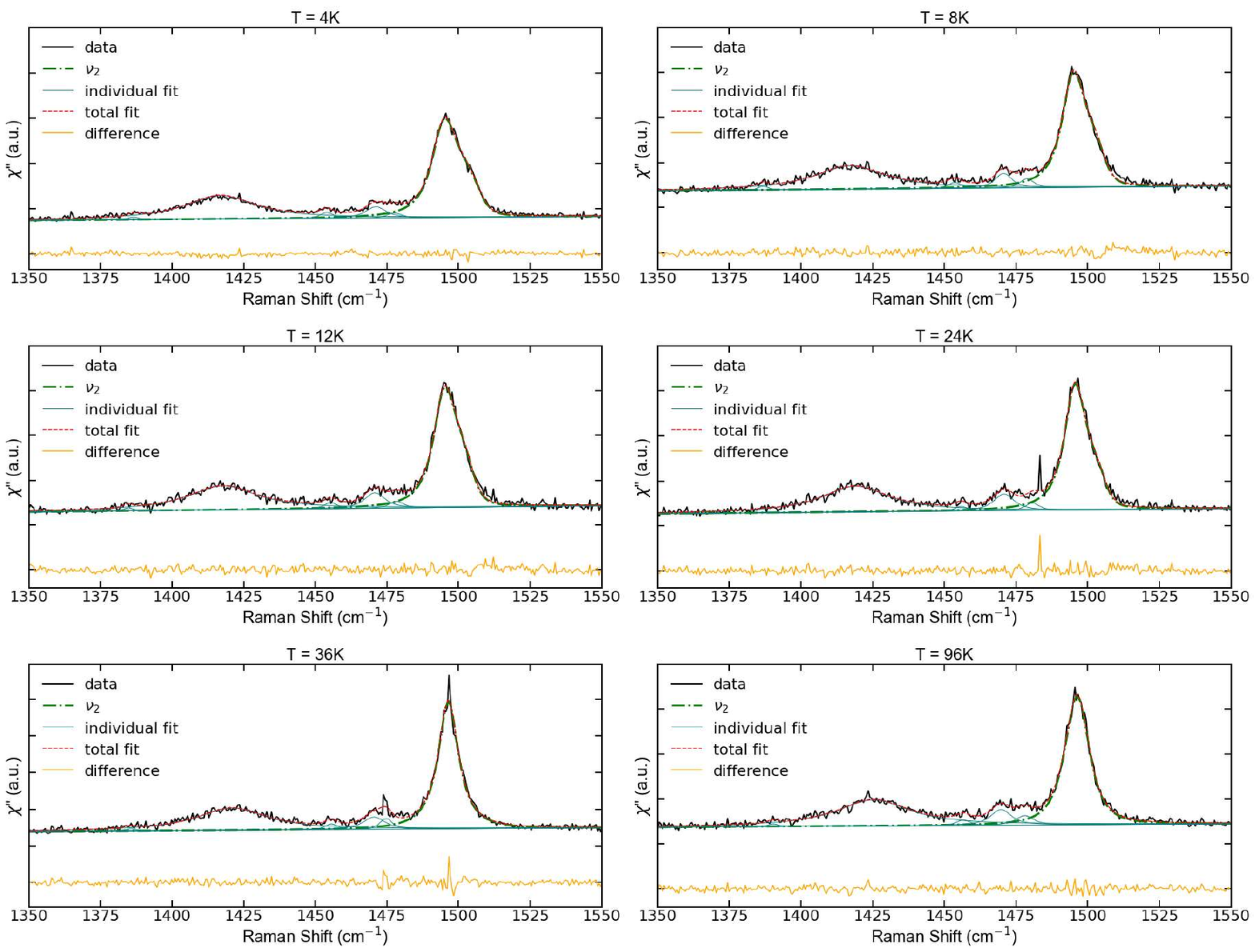}
        \caption{Fits of the fluctuating charge order in Sample 1}
        \label{SI_S1_tj}
\end{figure}
\pagebreak
\begin{figure}
    \centering
        \includegraphics[width=\linewidth]{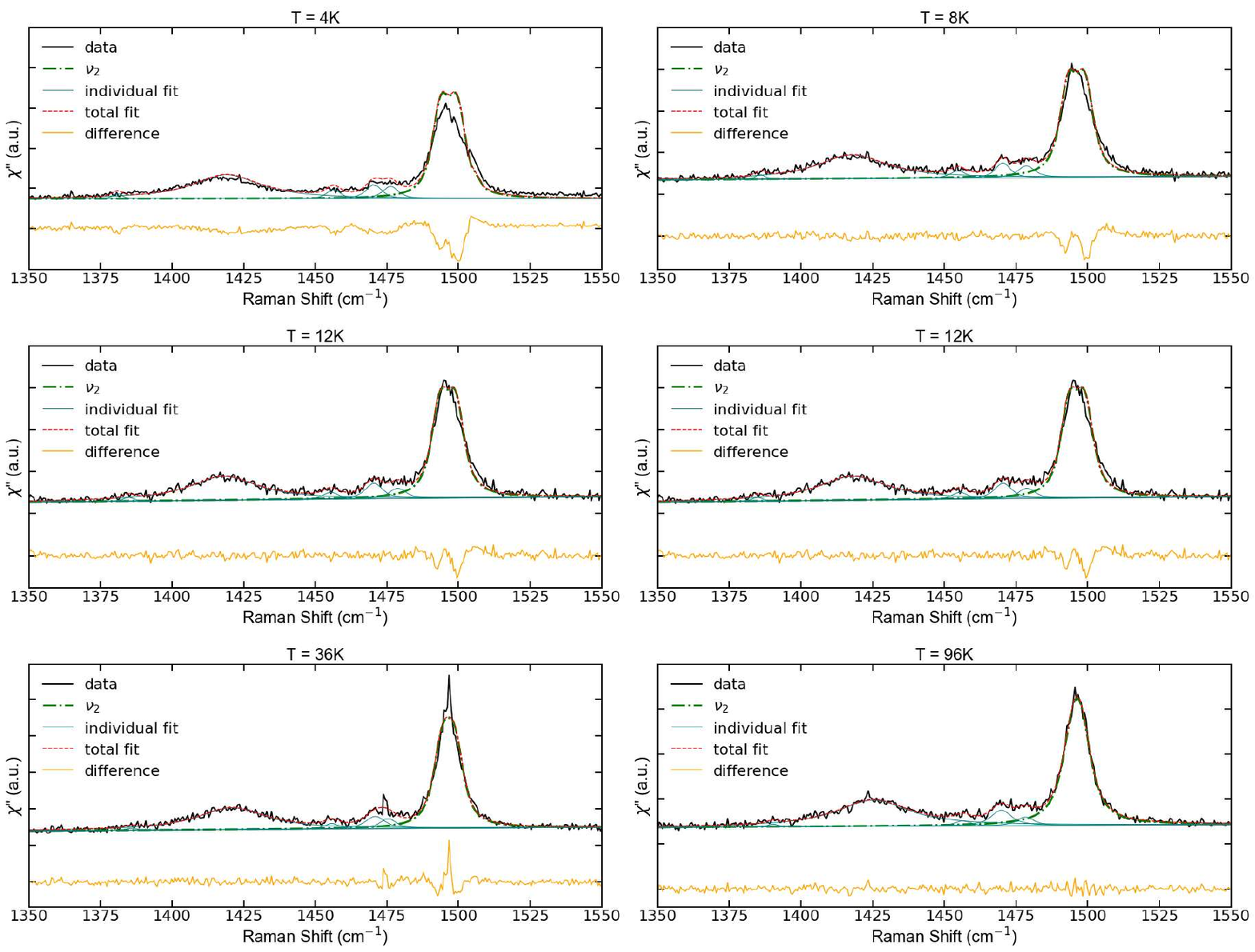}
        \caption{Fits of the static charge order in Sample 1}
        \label{SI_S1_static}
\end{figure}
\clearpage
\pagebreak
\begin{figure}
    \centering
        \includegraphics[width=\linewidth]{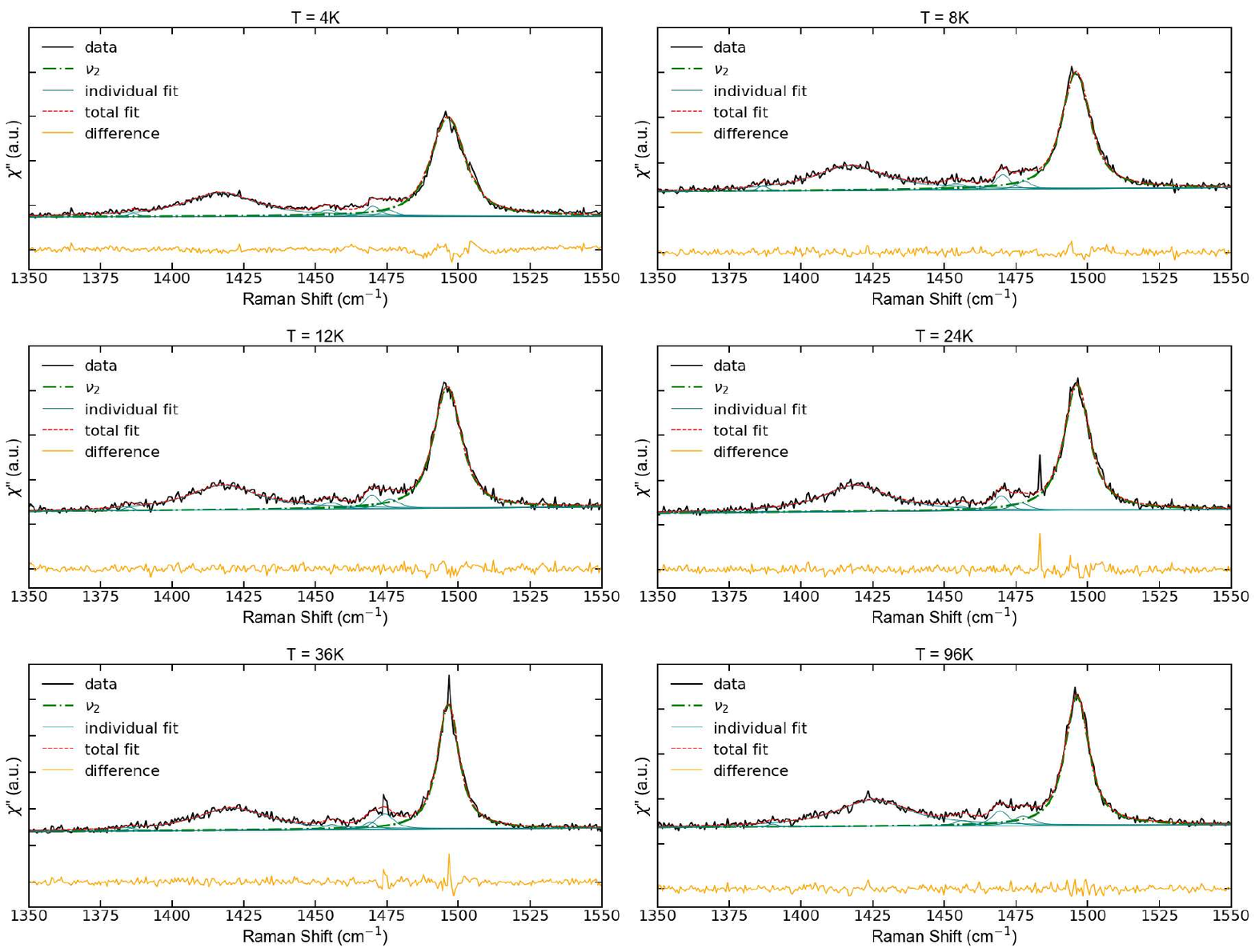}
        \caption{Fits of the charge glass in Sample 1}
        \label{SI_S1_gauss}
\end{figure}
\clearpage
\pagebreak
\begin{figure}
    \centering
        \includegraphics[width=\linewidth]{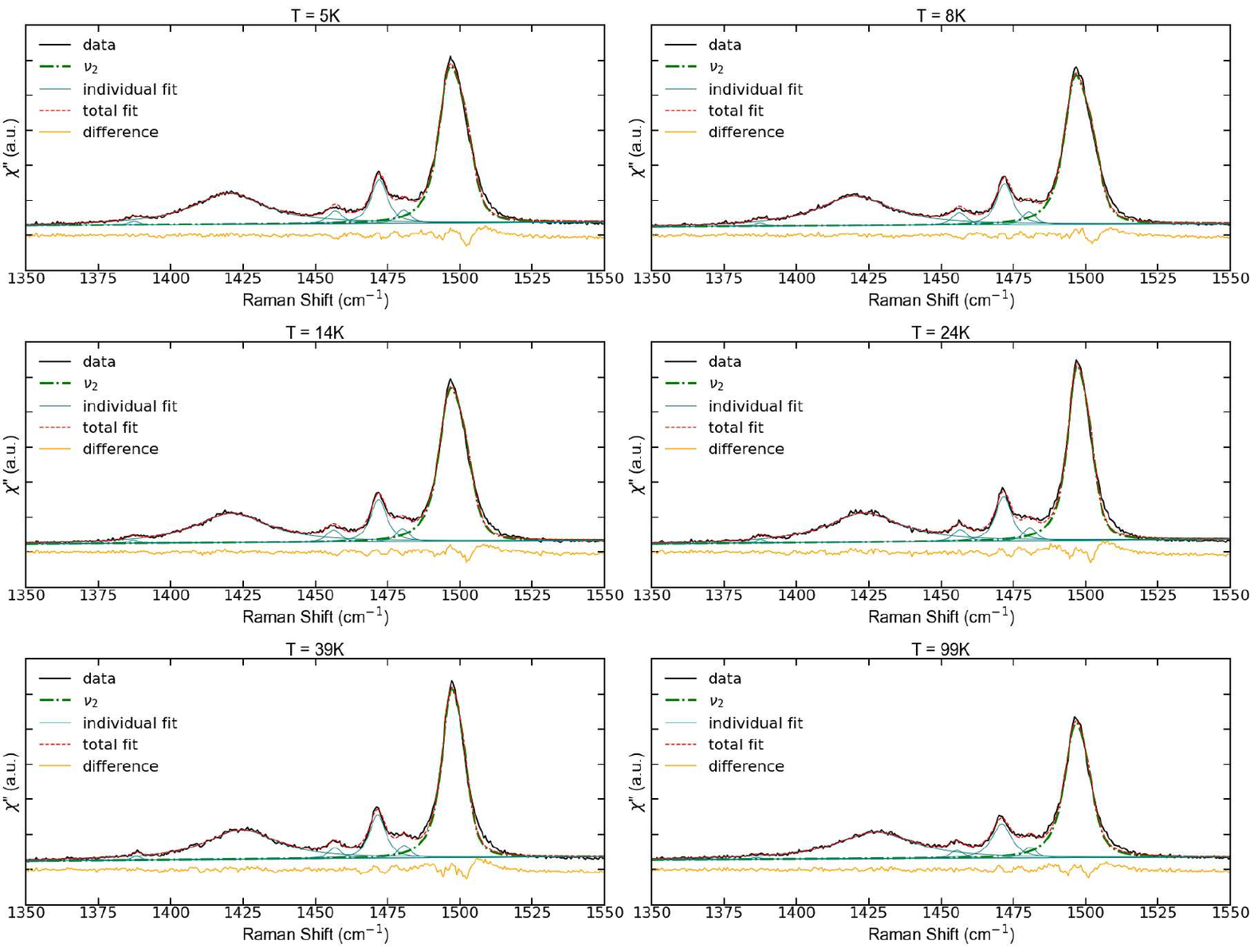}
        \caption{Fits of the fluctuating charge order in Sample 2}
        \label{SI_S2_tj}
\end{figure}
\pagebreak
\begin{figure}
    \centering
        \includegraphics[width=\linewidth]{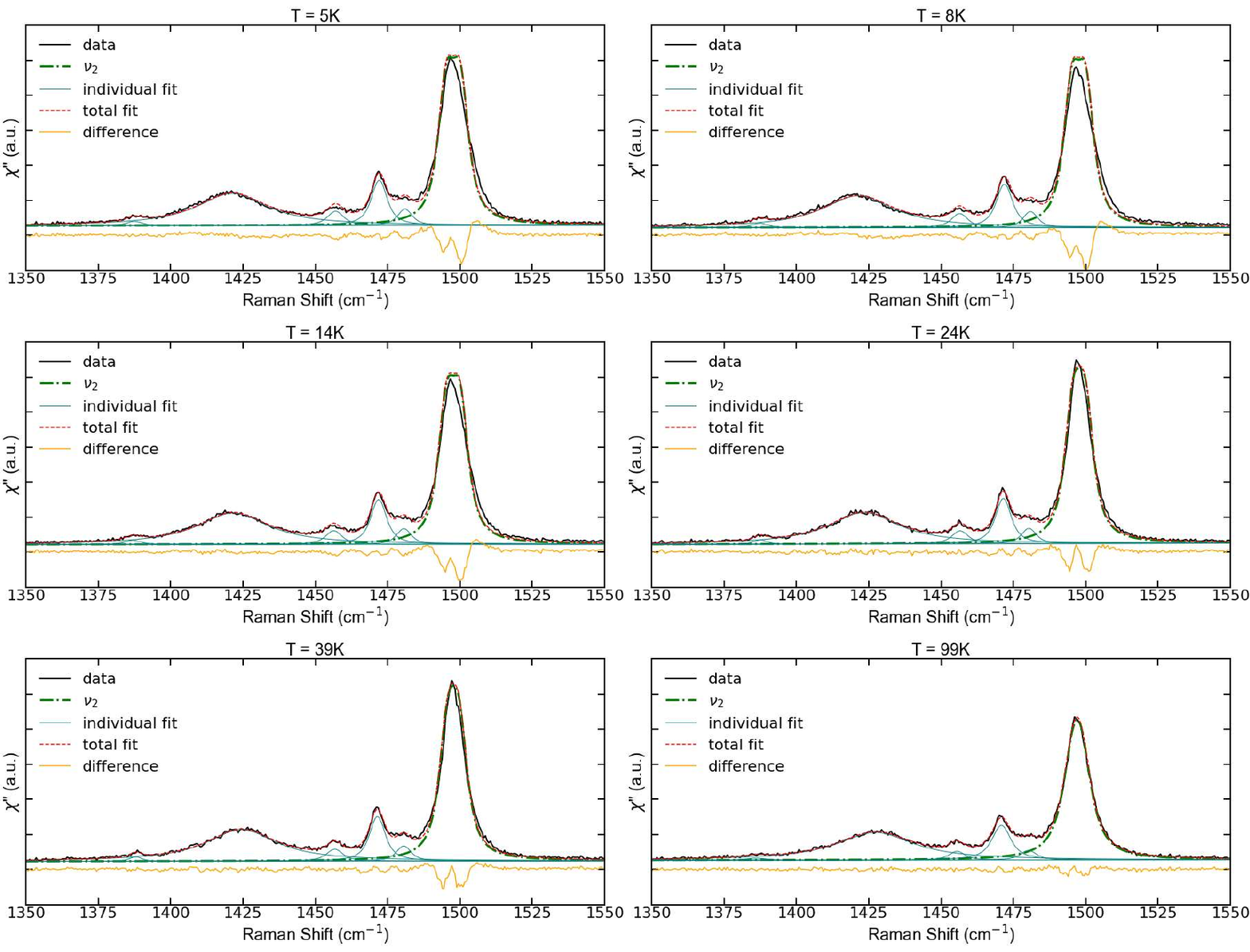}
        \caption{Fits of the static charge order in Sample 2}
        \label{SI_S2_static}
\end{figure}
\clearpage
\pagebreak
\begin{figure}
    \centering
        \includegraphics[width=\linewidth]{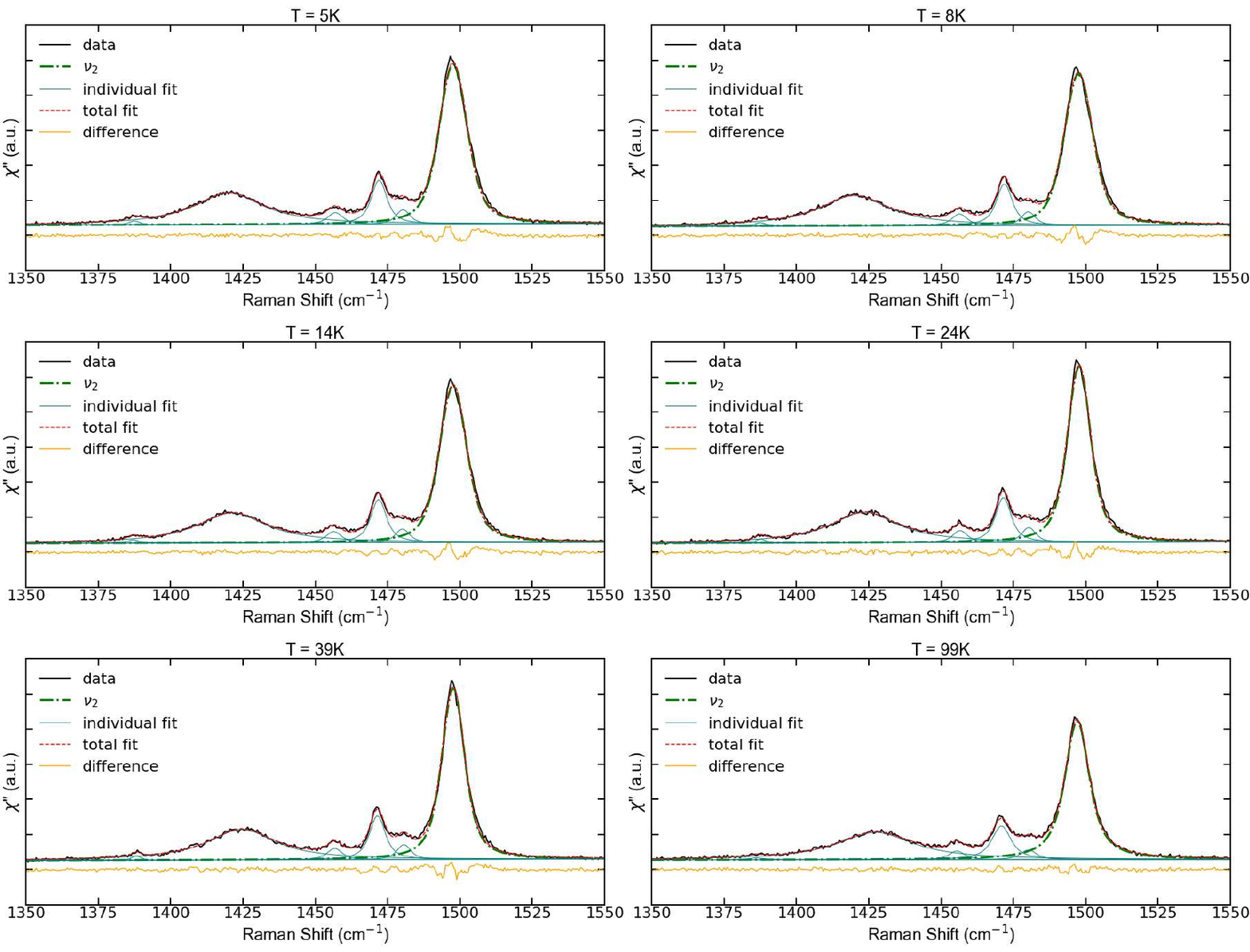}
        \caption{Fits of the charge glass in Sample 2}
        \label{SI_S2_gauss}
\end{figure}
\clearpage
\pagebreak
\begin{figure}
    \centering
        \includegraphics[width=\linewidth]{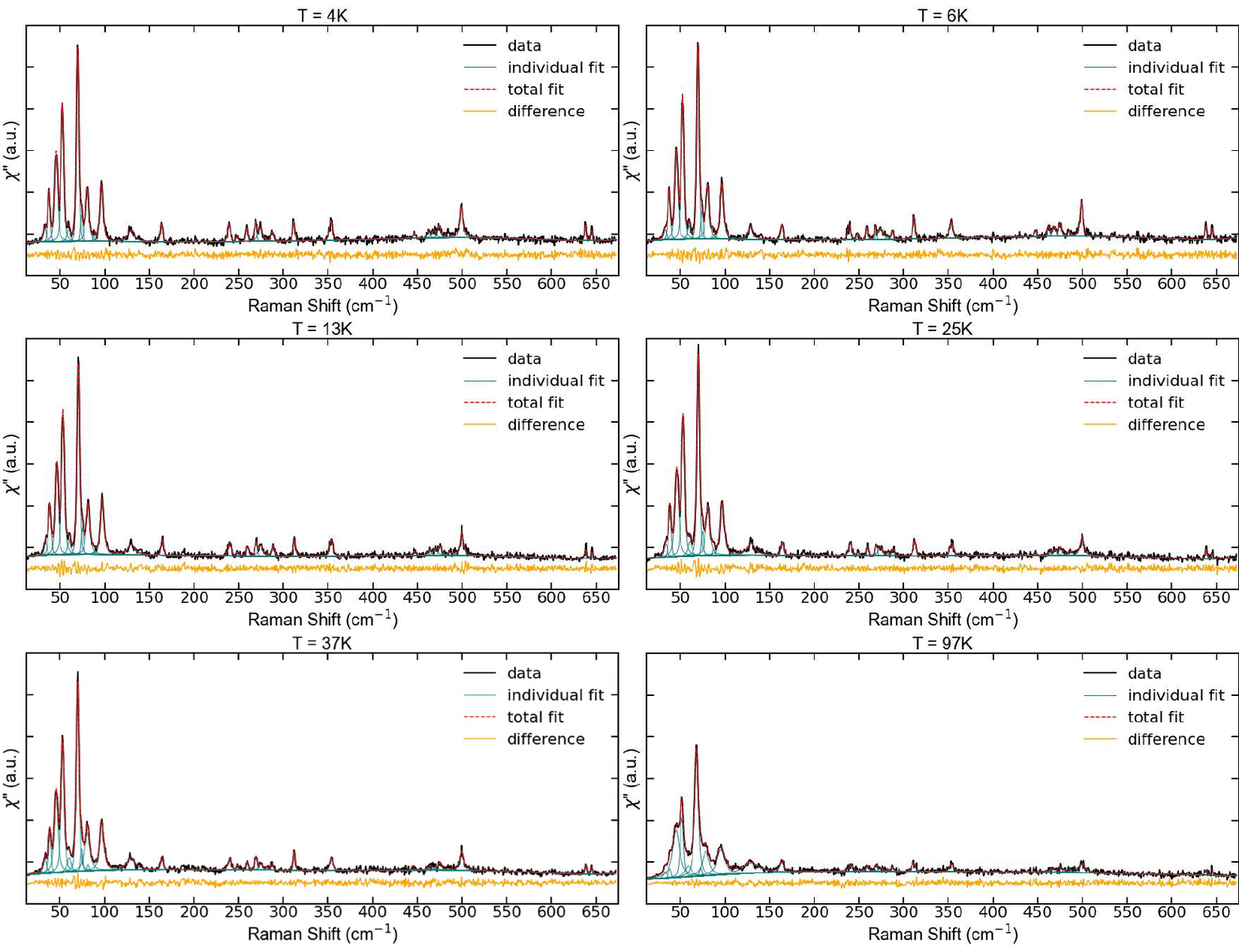}
        \caption{Fits in the bb polarization below 675~\cm}
        \label{SI_bb_fit}
\end{figure}
\clearpage
\pagebreak
\begin{figure}
    \centering
        \includegraphics[width=\linewidth]{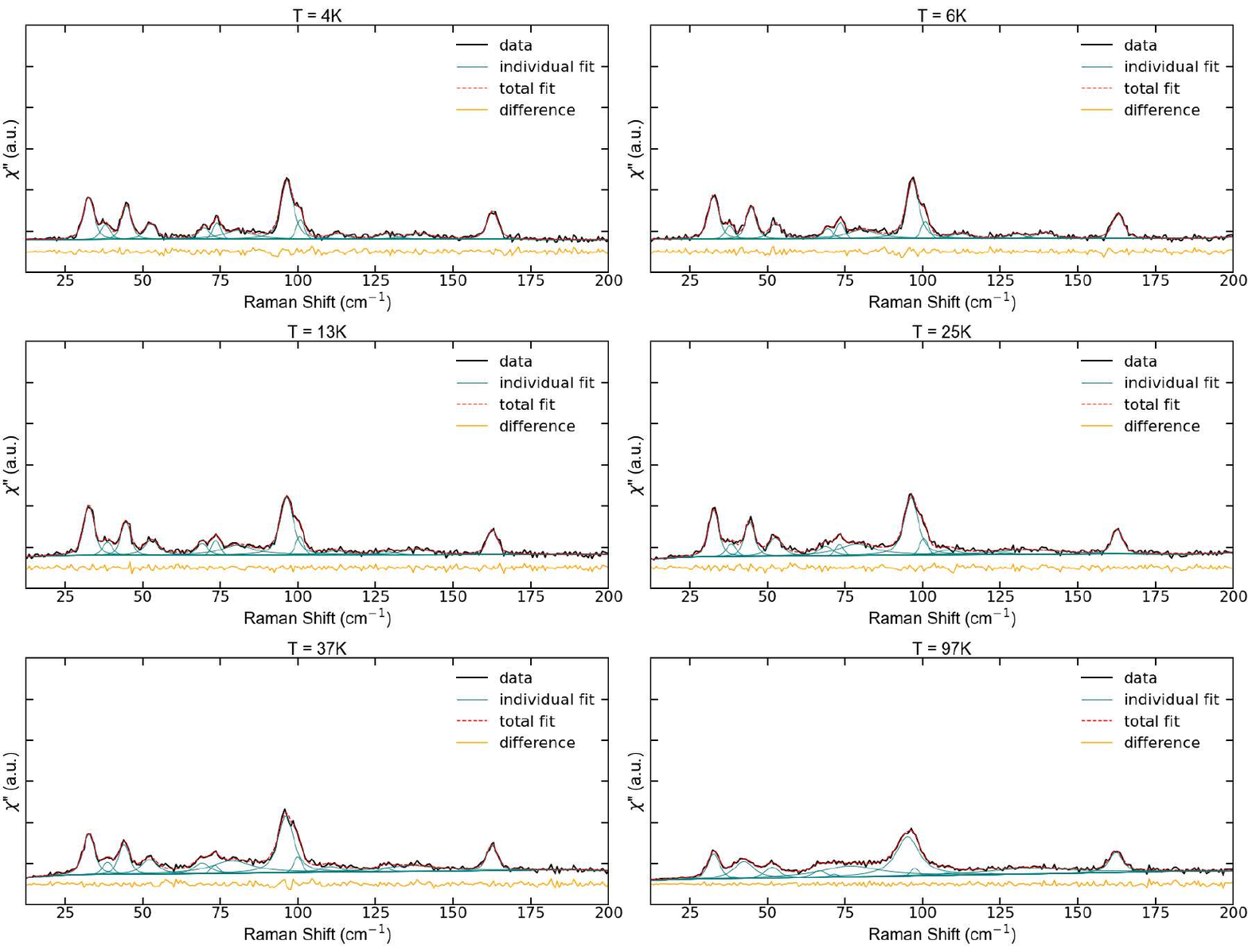}
        \caption{Fits in the bc polarization below 200~\cm}
        \label{SI_bc_fit}
\end{figure}
\clearpage
\pagebreak
\begin{figure}
    \centering
        \includegraphics[width=\linewidth]{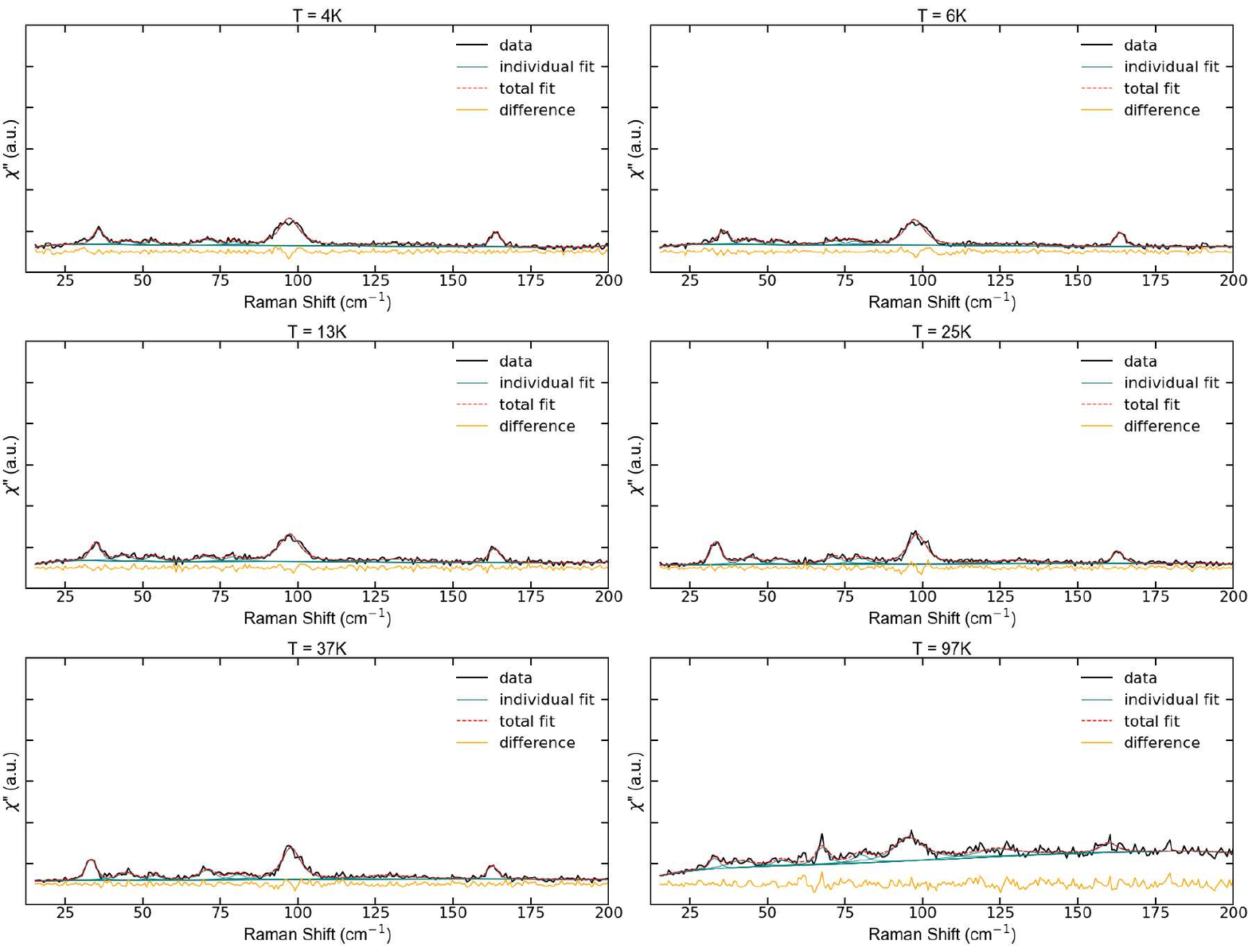}
        \caption{Fits in the cb polarization below 200~\cm}
        \label{SI_cb_fit}
\end{figure}
\clearpage
\pagebreak
\begin{figure}
    \centering
        \includegraphics[width=\linewidth]{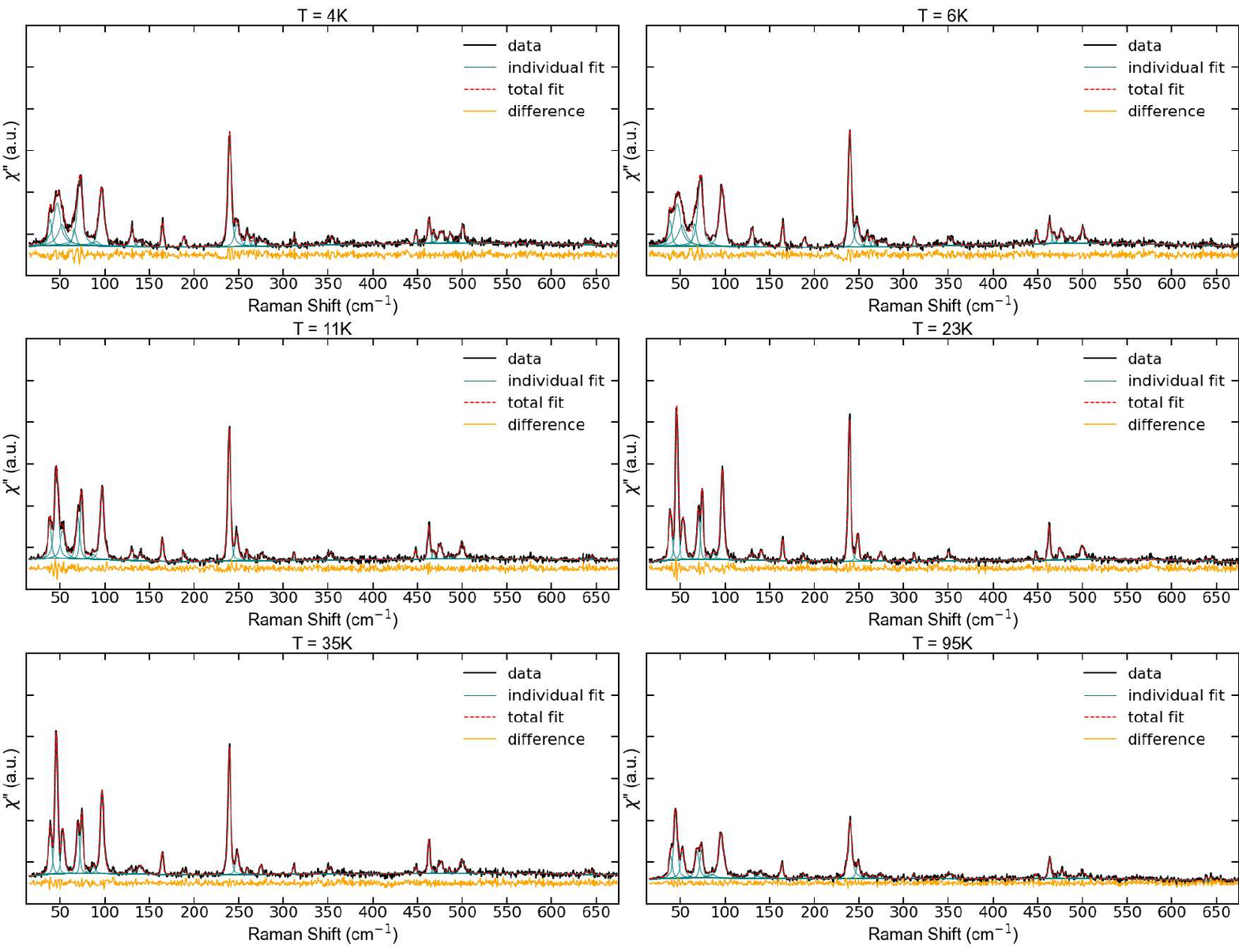}
        \caption{Fits in the cc polarization below 675~\cm}
        \label{SI_cc_fit}
\end{figure}
\clearpage
\pagebreak
\section*{Bibliography}

\bibliography{./SI.bbl}